\documentclass[10pt]{IEEEtran}
\usepackage{amsmath}
\usepackage[T1]{fontenc}
\usepackage{algorithm}
\usepackage{alphalph}
\usepackage{algorithmic}
\usepackage{listings}

\usepackage{nomencl}
\makenomenclature

\usepackage{amsmath}
\usepackage{graphicx}
\usepackage{color}
\usepackage[cmintegrals]{newtxmath}
\usepackage[caption=false,font=footnotesize]{subfig}
\ifCLASSOPTIONcompsoc
\usepackage[caption=false,font=normalsize,labelfont=sf,textfont=sf]{subfig}
\else
\usepackage[caption=false,font=footnotesize]{subfig}
\fi
\usepackage[colorlinks,citecolor=red,urlcolor=blue,bookmarks=false,hypertexnames=true]{hyperref} 
\usepackage{caption}
\usepackage{lipsum}
\usepackage{nameref}
\usepackage{varioref}
\usepackage{hyperref}
\usepackage{setspace}
\newtheorem{theorem}{Theorem}
\newtheorem{lemma}{Lemma}
\newtheorem{proposition}{Proposition}

\begin{document}
\title{\textcolor{black}{Novel Information-theoretic Game-theoretical Insights to} Broadcast\textcolor{black}{ing in} Internet-of-UAVs}
\author{Makan~Zamanipour,~\IEEEmembership{Member,~IEEE}
\thanks{Copyright (c) 2015 IEEE. Personal use of this material is permitted.
However, permission to use this material for any other purposes must be
obtained from the IEEE by sending a request to pubs-permissions@ieee.org. Makan Zamanipour is with Lahijan University, Shaghayegh Street, Po. Box 1616, Lahijan, 44131, Iran, makan.zamanipour.2015@ieee.org. }
}

\maketitle
\markboth{IEEE, VOL. XX, NO. XX, X 2021}%
{Shell \MakeLowercase{\textit{et al.}}: Bare Demo of IEEEtran.cls for Computer Society Journals}
\begin{abstract}
For the Internet-of-unmanned aerial vehicles (UAVs) some challenges in broadcasting and from new points of view are explored. In this paper, first, we investigate a single broadcast transceiver. From a control of noisy-channel viewpoint, we consider: (\textit{i}) Alice sends $\mathcal{X}$ to Bob as more \textcolor{black}{efficient} as possible while she wishes Bob not to get access to the private message $\mathcal{S}$ regarding the correlation between $\mathcal{S}$ and $\mathcal{X}$ $-$ i.e., Alice purposefully sends a \textit{turbulent-flow} of the information to Bob; and (\textit{ii}) where $\big (\Theta_1;\Theta_2 \big)$ is the control-action-pair which actualise a \textit{pursuit-Evasion}. We consider \textit{dissipativity} in our system due to the memory effect relating to the previous states. We thus propose a federated-learning based \textit{Blahut-Arimoto} algorithm while a 2-D \textit{dissipativity}-theoretic continuous-Mean-Field-Game (MFG) is proposed \textcolor{black}{with regard to} (w.r.t.) a joint probability-distribution-function (PDF) of the population distribution $-$ relating to a continuous-control-law. We also analyse what if Alice is owed to multiple Bobs in a multi-user scenario which we apply a bankruptcy based $3-$level nested game for.
\end{abstract}


\maketitle

\IEEEdisplaynontitleabstractindextext
\IEEEpeerreviewmaketitle

\section{Introduction} 

\IEEEPARstart{T}he information-theoretic bounds achieved through simultaneous resource use have a lot of priorities compared with the recent random access strategies. The main basis of the recent random access scenarios is to apply either collision avoidance $-$ such as ALOHA protocols $-$ or orthogonalization $-$ e.g.
time-division-multiple-access. In the former one, the decoder alarms an error if more than one transmitter is active in a time slot. Inversely, time slots are randomly selected by transmitters for the later one, while recourses being equally divided among the user.

Internet-of-unmanned aerial vehicles (UAVs) as the powerful Internet-of-Things (IoT) components are promising \cite{a1, a2, a3, a4, a5, a6}. In this area, UAV design technically requires to take into consideration a circumstance with an important set of uncertainties and dynamics. The quality of service relating to the sensing and communication without complete information is thus totally important \cite{a1, a2, a3, a4, a5, a6}.

In control over networked systems \cite{1, 2, 3, 4} and from an information-theoretic control-theoretic point of view, some brilliant activities have been performed so far. In \cite{2}, a probabilistic solution for the information constraints in a generic $NP-$hard optimisation problem was evaluated according to the \textit{source-and-channel separation}. In \cite{3}, the problem of controlling an unstable linear plant with Gaussian disturbances over an additive white Gaussian noise (AWGN) channel was theoretically investigated. By the usage of coding over a multiple channel, a single control action was mathematically realised. In \cite{4}, a CEO problem was evaluated for the multiple rounds of communication. A minimal mutual information was newly solved for the Gauss-Markov source observed via parallel AWGN channels. The authors in \cite{4} also bounded the rate-loss physically originated from the lack of communication among the observers. 

From an information theoretic game-theoretical point of view, although some rare work has been done e.g. \cite{5, 6, 7}, this dominant area of research is still open.
In \cite{5}, and for general alphabets, the saddle-point of the conditional Kullback-Leibler-Divergence to the conditional Re'nyi divergence was strongly evaluated. In \cite{6}, the information pattern of a mean-field-theoretical scheme was evaluated. In \cite{7}, a trade-off was realised in the context of ''how much fast and how much secure'' \textcolor{black}{with regard to} (w.r.t.) two essential limits in mean-field-games (MFGs). 

Some novel and traditional game-theoretical solutions have been done for the Internet-of-UAVs \cite{a1, a2, a3, a4, a5, a6}. In the aforementioned work, the metric was the information theoretic one \textit{age of information}.

In the context of the \textit{dissipativity} theory and from a fractional-order control theoretic point of view, the literature is also rich. In \cite{8} and for the first time, fractional-order congestion control oriented systems were introduced. Additionally, sufficient conditions were obtained as well in order to guarantee the occurrence of \textit{Hopf} bifurcations for fractional-order cases. Via a directed graph, \cite{9} explored the challenging issue of containment control in the fashion of fractional-order multi-agent systems.

From a control-theoretic point of view, the literature is also rich in the context of federated adapting. A variational inference procedure was proposed in \cite{10} for imitation learning of a control policy. The work in \cite{10} indeed was represented by a parametrised hierarchical algorithm which would invoke sub-procedures aimed at applying sub-tasks. In \cite{11}, how to control a system modeled by a probability density function (PDF) was explored which was performed by the calculation of a probabilistic control policy. A novel algorithm was proposed which was in terms of a data-driven iterative control in \cite{11}. How to optimally control of \textit{beliefs} was explored in \cite{12} while considering the role of a physical watermarking signal in fast-detection relating to an adversary in a scalar linear control system. The authors in \cite{12} indeed considered if one can replace the sensor measurements an arbitrary stationary signal generated by the adversaial.

\textit{Motivations and contributions:} In this paper, we are interested in responding to the following questions: \textit{What if we want to guarantee the privacy-utility trade-off \cite{13} for a problem of control over noisy communication channel? Is it possible to take into account a game in which the users are in terms of e.g. two groups of technique that a Chess Master has: (i) only one of which is supposed to be taught to his student; whereas (ii) he should be careful about the hostility probability of his student? How can we model the correlation between the two aforementioned groups of technique in the context of pursuit-evasion? How can we control of beliefs in the context of federated learning w.r.t. the correlation between the first group of techniques, i.e., the first group of agents, and the second specific and private one? What if the student’s hostility potential causes mixed-convolved actions? What if in a multi-user scenario, the sender owes to multiple receivers in the context of a bankruptcy?} With regard to the non-complete version of the literature, the expressed questions strongly motivate us to find an interesting solution, according to which our contributions are fundamentally described as follows. 
\begin{itemize}
\item \textcolor{red}{\textbf{(\textit{i})}} We evaluate the overall performance of the network from a joint continuous MFG-theoretical point of view. In relation to our MFG, the \textit{Fokker-Planck-Kolmogorov} (FPK) equation by which the PDF\footnote{Relating to a continuous-control-law.} of the population density is interpreted is two-dimensional, i.e., as a joint distribution relating to two groups of control actions. 

\item \textcolor{red}{\textbf{(\textit{ii})}} We also generalise the control-laws from a \textit{dissipativity theoretic} viewpoint by applying the generalised time-derivatives. 

\item \textcolor{red}{\textbf{(\textit{iii})}} A novel Blahut-Arimoto algorithm is subsequently proposed according to the federated learning principle in order to find the probability mass functions (PMFs) for the privacy-utility trade-off. This is also evaluated from a time-varying graph theoretic point of view. 

\item \textcolor{red}{\textbf{(\textit{iv})}} We also consider a multi-user fashion where we have multiple Bobs. We indeed consider a $3$-level nested game scenario where for the first type of the agents, that is, the decoders, we experience a bankruptcy. For the second class of $N+1-$player based game relating to the time instants, we apply a discrete mean-field-theoretical solution where the sensitivity about the bankruptcy-time-instant in terms of \textit{mean-field-limit} is considered. An alternating direction method of multipliers (ADMM) based mathematical expression is consequently traditionally proposed $-$ as common for the mean-field-limit issue. Instead of the ADMM based constraint, we define the third continuous MFG where the agents are in relation to phase transitions for the PMFs/PDFs. We do this in order to find a more adequate estimation of the probability spaces relating to the aforementioned ADMM based constraint, w.r.t. the \textit{Kuramoto} model. 
\end{itemize}

\textit{General notation:} The terms $(\cdot)_A$ or/and $(\cdot)^{(A)}$ stand for Alice while $(\cdot)_B$ or/and $(\cdot)^{(B)}$ stand for Bob. Additionally, $(\cdot)_1$ or/and $(\cdot)^{(1)}$ stand for the first game while $(\cdot)_2$ or/and $(\cdot)^{(2)}$ stand for the second one. Finally, $\setminus$, $Null(\cdot)$, $ \cdot|| \cdot$ and $dim(\cdot)$ are respectively stand for \textit{Set-minus}, Null, the Kullback-Leibler and dimension. Meanwhile, $\nabla$ stands for the Nabla symbol and also Bob as the controller is a map pair of $\bigg ( \Big (\hat{\mathcal{X}}^{(\ell)}_{k} \longmapsto \theta^{(\ell)}_{k,1} \Big); \Big(\hat{\mathcal{S}}^{(\ell)}_{k} \longmapsto \theta^{(\ell)}_{k,2} \Big) \bigg) $ as well.

\textbf{\textsc{Definition 1 \cite{14}.}} (\textit{i})\textit{ System $\sigma$ is controllable if for any pair
$(x_0, x_1) \in \mathbb{R}_n $ there exists a $T > 0$ and an input function $\theta(t) \in \mathbb{R}_m$ that
transfers the state $x(t)$ from $x_0$ at time $t = 0$ to $x_1$ at time $t = T$.} (\textit{ii}) \textit{Observability means that for any input function, the current state can be determined in finite time using only the outputs. In fact, the system is observable if the output w.r.t. the $x_0$ is equal to the output w.r.t. $x_t$, $\forall t>0$ and unobservable otherwise.} (\textit{iii}) \textit{A system is indeed detectable if we experience the output equating with zero for $x_0=0$, so, its steady state goes to zero, undetectable otherwise. }

\textbf{\textsc{Assumption 1. \textit{Anytime-}capacity \cite{1, 2, 3}:}} \textit{Throughout the paper, when we say rate, we refer to the principle of anytime-capacity. The f$-$anytime capacity $C_{\beta}(f)$ of a channel is the supremum of rates at which the channel can transmit data in the sense that (\textit{i}) the error probability is arbitrarily small, and (\textit{ii}) it decays at least as fast as $f(\cdot)$ does. $f(\mathscr{d}) > 0$ is any decreasing function of the delay $\mathscr{d}$ e.g. $f(\mathscr{d}) = 2 ^{-\beta \mathscr{d}}$. Indeed, anytime capacity is the upper-bound of the error-free capacity being equated when $\beta \rightarrow \infty$ and it is the lower-bound of the Shannon-capacity being equated when $\beta \rightarrow 0$.\footnote{As a case in point: $C_{\alpha}(f) \ge 1 - log_2 (1 + \mathscr{e}) - \beta$ holds for a Binary Erasure Channel with the erasure probability of $e$.} The logic behind of the anytime capacity is the fact that the encoder-decoder pair must be anytime, i.e., timely synchronisable or real-time.}

\textbf{\textsc{Remark 1:}} \textit{Instead of the time-derivative $\frac{d}{dk}$, one should use the generalised fractional-order time-derivative $-$ see e.g. \cite{8, 9} $-$ $\mathcal{D}_k^{(\alpha)}=\frac{1}{\Gamma (1-\alpha)} \frac{d}{dk} \int_0^k f(\kappa) \big(k-\kappa \big)^{\alpha} d\kappa, \alpha \in [0,1]$, as well as the fractional-order gradient operator $\nabla_k^{(\alpha)}=\frac{1}{\Gamma (1-\alpha)} \int_0^k \nabla_k f(\kappa) \big(k-\kappa \big)^{\alpha} d\kappa, \alpha \in [0,1]$ where $\Gamma(x)=\int_0^{\infty} x^{z-1} e^{-x}dx$. The reason is justified as follows. It should be noted that, the global system may act as a non-linear or non-equilibrium thermodynamic steady-state even if there exist some local equilibria. In other words, the global system may be time-irreversible and the global dynamics of the process don’t remain \textit{well-defined} $-$ i.e., it no longer holds $-$ when the sequence of the global time-states, i.e., global dynamic equations are irreversible. Indeed, the global system may consist of mixed-convolved actions e.g. due to the Bob’s falsification, the time-varying mean and variance for the noise etc. Therefore, since the system may be a dissipative\footnote{See the \textit{dissipativity theory}. For short, a dissipative system such as a tornado is thermodynamically open, i.e., the one operating far from thermodynamic equilibria. In other words, a dissipative system lies theoretically in a dynamical fashion with a reproducible steady-state $-$ in contrast to conservative ones. One example would be the Hopf bifurcation which emphatically imposes that dynamical systems can be physically decomposed into two absolute parts: (\textit{i}) a dissipative one; and (\textit{ii}) a conservative one.} one, a convolution of the derivative of the main function with a power function $-$ which stands for the time-oscillations and fluctuations in relation to the memory effect over the previous states $-$ is required. In order to take into account the point expressed here, we should formulate the time-derivatives with the generalised fractional-order ones.}

\subsection{Organisation}
The rest of the paper is organised as follows. The system set-up and our main results are given in Sections II and III. Subsequently, the evaluation of the framework and conclusions are given in Sections IV and V.

\section{System model and problem formulation}
In this section, we describe the system model from both information-theoretic and control-theoretic viewpoints, subsequently, we formulate the basis of our problem. 

\textcolor{black}{Initially speaking, let us call $3$ random variables ${\mathcal{S}}$, ${\mathcal{X}}$ and ${\mathcal{Y}}$ which are i.i.d discrete memoryless variables respectively with the PMFs $\mathcal{P}_{\mathcal{S}}$, $\mathcal{P}_{\mathcal{X}}$ and $\mathcal{P}_{\mathcal{Y}}$}. From an information-theory point of view assume that a sender named Alice has some private data denoted by the random variable $\mathcal{S} \in S$ which is correlated with some non-private data $\mathcal{X} \in X$. Alice is supposed to share $\mathcal{X}$ with an analyst named Bob. However, due to the correlation between $\mathcal{X}$ and $\mathcal{S}$ which is captured by the joint distribution $\mathcal{P}_{\mathcal{S},\mathcal{X}}$, Bob may be able to draw some inference on the private data $\mathcal{S}$. Alice consequently decides to, instead of $\mathcal{X}$, release a distorted version of $\mathcal{X}$ defined by $\mathcal{Y} \in Y$ in order to alleviate the inference threat over $\mathcal{S}$ logically acquirable from the observation of $\mathcal{X}$. The distorted data $\mathcal{Y}$ is generated by passing through the following privacy mapping, i.e., the conditional distribution $\mathcal{P}_{\mathcal{Y}|\mathcal{X}}$. It should be noted that, in fact, Bob may also be able to act as an adversary by using $\mathcal{Y}$ to illegitimately infer the private data set $\mathcal{S}$, even though he is a legitimate recipient of the data set $\mathcal{Y}$. Therefore, the privacy mapping should be designed in the sense that we can be assured about a reduction to the inference threat on the private set $\mathcal{S}$ as follows: while preserving the utility of $\mathcal{Y}$ by maintaining the correlation, i.e., dependency between $\mathcal{Y}$ and $\mathcal{X}$, we aim at alleviating the dependency between $\mathcal{S}$ and $\mathcal{Y}$. This kind of two-fold information-theoretic goal balances a trade-off between utility and privacy \textcolor{black}{\cite{13}}. As also obvious, the Markov chain $\mathcal{S} \rightarrow \mathcal{X}\rightarrow \mathcal{Y}$ holds \textcolor{black}{\cite{13}}. 

Let us go in datails. For the $k$-the time instant where $k \in [0,\mathcal{K}]$ including $k \rightarrow \infty$ $-$ which declares that we use the \textit{any-time capacity} principle (see \textit{Assumption 1}),  $-$ Alice observes the source-symbol sets $-$ sequences $-$ $\mathcal{S}^{(\ell)}_k=\{ s_k^{(\ell)} \}^{\mathcal{L}_1}_{\ell=1}$ and $\mathcal{X}^{(\ell)}_k=\{ x_k^{(\ell)} \}^{\mathcal{L}_2}_{\ell=1}$ where Bob observes $\mathcal{Y}^{(\ell)}_{k}=\{ y_{k}^{(\ell)} \}^{\mathcal{L}_3}_{\ell=1}$ \textcolor{black}{while} $\{ \cdot \}^{\big(\ell \in \left \{  1,\cdots,\mathcal{L}_i, i \in \{ 1,2,3 \}  \right \}  \big)}$ stands literally for the $\ell$-th source-symbol per block belonging to the source-symbol set of size $1$-by-$\mathcal{L}_i, i \in \{ 1,2,3 \}$. In fact, we principally see a simple scheme where: (\textit{i}) the Alice-encoder follows the Borel measurable map $f^{(\ell)}_{x,k}: \;\mathcal{X}^{(\ell)}_{k}  \longmapsto \mathcal{M}^{(x,k)}=\{ 1, 2,\cdots, 2^{\ell \mathcal{R}^{(\ell)}_{x,k}} \}$, while $\mathcal{R}^{(\ell)}_{x,k}$ is the rate of Alice; moreover, (\textit{ii}) the Bob-decoder legitimately follows the Borel measurable map $g^{(\ell)}_{k}: \; \{ 1,\cdots, 2^{\ell \mathcal{R}^{(\ell)}_{x,k}} \}  \longmapsto \mathcal{X}^{(\ell)}_{k}$ after reception via the channel and illegitimately the map $\varphi \big ( \mathcal{S}\big )\longrightarrow \mathcal{S}$. The term $\cdot \longmapsto \cdot$ also stands for the source-encoding/decoding process.
 Therefore, the probability of reconstruction-error is $\mathcal{P} \bigg( \hat{\mathcal{X}}^{(\ell)}_{k} \neq \mathcal{X}^{(\ell)}_{k} \bigg)=\sum_{\ell }  \sum_{k } \mathbb{E} \big\{   {\hat{\mathcal{X}}^{(\ell)}_{k} } \neq \mathcal{X}^{(\ell)}_{k}   \big\}$ where $\hat{\mathcal{X}}^{(\ell)}_{k}$ is the reconstructed version of $\mathcal{X}^{(\ell)}_{k} $ done by the Bob-decoder. 

Additionally, in order to guarantee a trade-off between utility and privacy as discussed above, we should let the following be satisfied: $\mathop{{\rm min}}\limits_{\mathcal{P}{\big(\mathcal{Y}|\mathcal{X}} \big)} {\rm \; } \mathcal{I} \big(\mathcal{Y},\mathcal{S}\big) \;\; s.t. \;\; \frac{1}{dim \bigg(  Null  \Big( \mathcal{I}\big(\mathcal{X},{\mathcal{Y}}\big) \Big) \bigg)}=dim \bigg(  Full  \Big( \mathcal{I}\big(\mathcal{X},{\mathcal{Y}}\big) \Big) \bigg) \longrightarrow \epsilon^{+}$ or equivalently\footnote{The distortion $\frac{1}{dim \bigg(  Null  \Big( \mathcal{I}\big(\mathcal{X},{\mathcal{Y}}\big) \Big) \bigg)}$ can be re-written as $\mathbb{E} \left \{            \mathcal{P}  \big(\mathcal{S}  |\mathcal{X}   \big) || \mathcal{S}   \big ( \mathcal{S}  |\mathcal{Y}  \big)  \right \}  =\sum_{x,y}   \mathcal{P}  \big(\mathcal{X}  |\mathcal{Y}   \big) \sum_{s}    \mathcal{P}  \big(\mathcal{S}  |\mathcal{X}   \big) log   \frac{ \mathcal{P}  \big(\mathcal{S}  |\mathcal{X}   \big)}{ \mathcal{P}  \big(\mathcal{S}  |\mathcal{Y}   \big)}  =\sum{x,y,s}  \mathcal{P}  \big(\mathcal{X} ,\mathcal{Y}  ,\mathcal{S}    \big) log   \frac{ \mathcal{P}  \big(\mathcal{S}  |\mathcal{X}   \big)}{ \mathcal{P}  \big(\mathcal{S}  |\mathcal{Y}   \big)} =\mathcal{I} \big(   \mathcal{X},\mathcal{S} \big) -\mathcal{I} \big(   \mathcal{Y},\mathcal{S} \big)$.} $\mathop{{\rm min}}\limits_{\mathcal{P}{\big(\mathcal{Y}|\mathcal{X}} \big)} {\rm \; } \mathcal{I} \big(\mathcal{Y},\mathcal{S}\big) \;\; s.t. \;\; \mathcal{I} \big(   \mathcal{X},\mathcal{S} \big) -\mathcal{I} \big(   \mathcal{Y},\mathcal{S} \big) \longrightarrow \epsilon^{+}$, that is, in the context of $\mathop{{\rm max}}\limits_{(\cdot)} {\rm \; } Privacy \;\; s.t. \;\; Utility$. The term $\epsilon^{+}$ is also the non-zero distortion threshold.

From a control-theory point of view Bob as the controller is a map pair of $\bigg ( \Big (\hat{\mathcal{X}}^{(\ell)}_{k} \longmapsto \theta^{(\ell)}_{k,1} \Big); \Big(\hat{\mathcal{S}}^{(\ell)}_{k} \longmapsto \theta^{(\ell)}_{k,2}  \Big) \bigg) $, since as discussed before, Bob may also be able to illegitimately infer $\mathcal{S}$. In other words, Bob observes $\mathcal{Y}$ which is: (\textit{i}) a distorted version of $\mathcal{X}$; as well as (\textit{ii}) a non-linear function $\varphi \big(  \mathcal{S}  \big)$ due to the correlation between $\mathcal{X}$ ans $\mathcal{S}$.

\section{Main results}
\textsc{\textbf{Objective.}} \textit{Our task is to examine if we can construct encoders and controllers such that the closed loop system is stable and detectable, i.e., if the triple $(\mathcal{A}, \mathcal{B}_1,\mathcal{B}_2)$ is controllable-and-stabilisable and if the pair $(\mathcal{A}, \mathcal{C})$ is observable-and-detectable. }
          
Now, let us do the following exemplification. Take into account a game in which the users are in terms of e.g. two groups of technique that a Chess Master has: (\textit{i}) the first group is supposed to be taught to his student; and (ii) the second group is the one which is not, since he should be careful about the hostility probability of his student. He is fully aware of the correlation between the two groups of techniques $-$ i.e., the first group of agents and the second specific and private one $-$ according to which the student may be able to infer the second ones. Additionally, there is albeit an interaction among the agents in each group of techniques as obvious. 

\textsc{\textbf{Remark 2:}} \textit{The correlation between the two aforementioned groups of technique models a pursuit-evasion.}

\textsc{\textbf{Remark 3:}} \textit{Although in the game between Alice and Bob, the \textit{Leader} is Alice at the first glance, according to the fact of the hostility potential of Bob, Alice may also be able to be the \textit{Follower}. }

\subsection{Our game when Bob may illegitimately act: A generic point of view}
\begin{lemma} \label{L2} \textit{From the $\mathcal{I} \big( \mathcal{Y},\mathcal{S} \big)$ point of view, Bob and Alice may experience a non-cooperative non-zero-sum game, although both of them agree on $\mathcal{I} \big(   \mathcal{X},\mathcal{Y} \big) \neq 0$, where Alice and Bob follow $\mathcal{I} \big(   \mathcal{X}^{(\ell^{\prime})}_{k^{\prime}},\mathcal{S}^{(\ell^{\prime\prime} )}_{k^{\prime\prime} } \big) \neq 0$ and $ \mathcal{I} \big(   \mathcal{Y}^{(\ell^{\prime})}_{k^{\prime}},\mathcal{S}^{(\ell^{\prime\prime} )}_{k^{\prime\prime}}  \big) \neq 0$, respectively.}\end{lemma}

\textbf{\textsc{Proof:}} For respectively Alice and Bob, assume that we have a game where $\mathcal{S}^{(\ell)}_k=\{s^{(\ell)}_k\}^{\mathcal{L}_1}_{\ell=1}$ and $\mathcal{Y}^{(\ell)}_k=\{y^{(\ell)}_k\}^{\mathcal{L}_3}_{\ell=1}$ are the agents (see e.g. \cite{6, 7}). For each group of agents, let us select the ones whose information-gain is non-zero as follows as the major ones, i.e., the Headers: (\textit{i}) at Alice $\mathcal{I} \big(   \mathcal{X}^{(\ell^{\prime})}_{k^{\prime}},\mathcal{S}^{(\ell^{\prime\prime} )}_{k^{\prime\prime} } \big) \neq 0$; and (\textit{ii}) at Bob $ \mathcal{I} \big(   \mathcal{Y}^{(\ell^{\prime})}_{k^{\prime}},\mathcal{S}^{(\ell^{\prime\prime} )}_{k^{\prime\prime}}  \big) \neq 0$, where either $k^{\prime}=k^{\prime\prime}$ or $k^{\prime} \neq k^{\prime\prime}$ or/and $\ell^{\prime}=\ell^{\prime\prime}$ or $\ell^{\prime} \neq \ell^{\prime\prime}$. Here, w.r.t. $\mathcal{I} \big( \mathcal{X}^{(\ell^{\prime})}_{k^{\prime}},\mathcal{S}^{(\ell^{\prime\prime} )}_{k^{\prime\prime} } \big) \neq 0$, Alice wishes 
\begin{equation*}
\begin{split}
\; \mathcal{I} \big(   \mathcal{Y}^{(\ell^{\prime})}_{k^{\prime}},\mathcal{S}^{(\ell^{\prime\prime} )}_{k^{\prime\prime}}  \big) < \partial^{(k_0,\ell_0)}_3,
\end{split}
\end{equation*}
to be held, whereas this is the inverse wish of Bob since Bob wishes 
\begin{equation*}
\begin{split}
\; \mathcal{I} \big(   \mathcal{Y}^{(\ell^{\prime})}_{k^{\prime}},\mathcal{S}^{(\ell^{\prime\prime} )}_{k^{\prime\prime}}  \big) > \partial^{(k_0,\ell_0)}_4, k_0 \in \{ k^{\prime} ,k^{\prime\prime} \},\ell_0  \in  \{ \ell^{\prime} ,\ell^{\prime\prime} \},
\end{split}
\end{equation*}
where $\partial_3$ and $\partial_4$, $\partial_4 >> \partial_3$, are some thresholds.

This completes the proof.$\; \; \; \blacksquare$

\begin{lemma} \label{L3} \textit{From the $\mathcal{I} \big( \mathcal{X},\mathcal{Y} \big)$ point of view, Bob and Alice may experience a non-cooperative non-zero-sum game, w.r.t. $\mathcal{I} \big( \mathcal{X},\mathcal{Y} \big) \neq 0$.}\end{lemma}

\textbf{\textsc{Proof:}} Although both Alice and Bob agree on $\mathcal{I} \big( \mathcal{X},\mathcal{Y} \big) \neq 0$, Alice wishes both $\mathcal{I} \big( \mathcal{X},\mathcal{Y} \big) > 0$\footnote{I.e.: The \textit{Radon-Nikodym} derivative through a data-dependent partition of the observation
space, as the \textit{projection} is non-zero and $\mathcal{X} \bot \mathcal{Y}$ does not hold.} and $\mathcal{I} \big( \mathcal{X},\mathcal{Y} \big) < \partial_2$ to be held, whereas Bob wishes $\mathcal{I} \big( \mathcal{X},\mathcal{Y} \big) \longrightarrow \infty$ to be held. 

This completes the proof.$\; \; \; \blacksquare$

\subsection{Our first MFG: $ \mathcal{I} \big(   \mathcal{Y}_{k},\mathcal{S}_{k}  \big) $}
\begin{proposition} \label{P1}
 \textit{Our first MFG in relation to the term $ \mathcal{I} \big(   \mathcal{Y}_{k},\mathcal{S}_{k}  \big) $ is presented as follows:} 
\begin{itemize}
\item \textcolor{red}{\textbf{(\textit{i})}} \textit{\textsc{Control-Law}: According to Lemma 1, one can write the following control-laws: }
\begin{itemize}
\item \textit{(i) for Alice as}
\begin{equation*}
\begin{split}
\; \mathcal{D}_k^{(\alpha)}   \left \{ \mathcal{I} \big(   \mathcal{Y}_{k},\mathcal{S}_{k}  \big)  \right \}  = \;\;\;\;\;\;\;\;\;\;\;\;\;\;\;\;\;\;\;\;\;\;\;\;\;\;\;\;\;\;\;\;\;\;\;\;\;\;\;\;\;\;\;\;\;\;\;\; \\- \Phi_1 \Big (\mathcal{P} \big(   \mathcal{Y}_{k}|\mathcal{X}_{k}  \big)  ; \Phi^{\prime}_1 (k;\alpha);k \Big) + \mathcal{W}^{(k)}_1;
\end{split}
\end{equation*}
and
\item \textit{(ii) for Bob as}
\begin{equation*}
\begin{split}
\; \mathcal{D}_k^{(\alpha)}   \left \{ \;  \mathcal{I} \big(   \mathcal{Y}_{k},\mathcal{S}_{k}  \big) \right \} \propto  \mathcal{D}_k^{(\alpha)}  \left \{ \;  \mathcal{I} \big(   \hat{\mathcal{X}}_{k},\mathcal{S}_{k}  \big) \right \}  =
\;\;\;\;\;\;\;\;\;\;\; \\
 + \Phi_2 \Big (\mathcal{P} \big(   \mathcal{Y}_{k}|\mathcal{X}_{k}  \big)  ; \Phi^{\prime}_2 (k;\alpha);k \Big) + \mathcal{W}^{(k)}_2,
\end{split}
\end{equation*}
where $\mathcal{W}^{(k)}_i,i \in \{ 1,2 \}$ is a time-varying random walk Winner process which stands for the relative gradients, and we assume $|\Phi^{(k)}_i |>0,i\in \{1,2\}$ without loss of generality\footnote{In order to preserve the type of the law’s conservation/non-conservation.}.
\end{itemize}

\item \textcolor{red}{\textbf{(\textit{ii})}} \textit{\textsc{Value function:} The value function is written:}
\begin{itemize}
 \item \textit{(i) for Alice as }
\begin{equation*}
\begin{split}
\; \mathcal{V} _{A,1}\Big(k;\mathcal{P}{\big(\mathcal{Y}|\mathcal{X}} \big) \Big)\triangleq \mathop{{\rm min}}\limits_{\mathcal{P}{\big(\mathcal{Y}|\mathcal{X}} \big)}\mathbb{E} \left \{ \int_0^\mathcal{K}  \mathcal{I} \big(   \mathcal{Y}_{k},\mathcal{S}_{k}  \big) dk \right \};
\end{split}
\end{equation*}
and
\item \textit{(ii) for Bob as }
\begin{equation*}
\begin{split}
\; \mathcal{V} _{B,1}\Big(k;\mathcal{P}{\big(\mathcal{Y}|\mathcal{X}} \big) \Big)\triangleq\mathop{{\rm max}}\limits_{\mathcal{P}{\big(\mathcal{Y}|\mathcal{X}} \big)}\mathbb{E} \left \{ \int_0^\mathcal{K} \mathcal{I} \big(   \hat{\mathcal{X}}_{k},\mathcal{S}_{k}  \big)dk \right \}.
\end{split}
\end{equation*}
\end{itemize}

\item \textcolor{red}{\textbf{(\textit{iii})}} \textit{\textsc{HJB:} The Hamiltonian-Jacobi-Bellman (HJB) equation is written: }
\begin{itemize}
\item \textit{(i) for Alice as }
\begin{equation*}
\begin{split}
\; HJB_{A,1}\triangleq\; \mathcal{D}_k^{(\alpha)}   \left \{ \mathcal{V}  _{A,1}\Big(k;\mathcal{P}{\big(\mathcal{Y}|\mathcal{X}} \big) \Big)\right \}\;\;\;\;\;\;\;\;\;\;\;\;\;\;\;\;\;\;\;\;\;\;\;\;\;\;\;\; \;\;\;\;\;\\+ \mathop{{\rm min}}\limits_{\mathcal{P}{\big(\mathcal{Y}|\mathcal{X}} \big)} \rm \;  \mathcal{I} \big(   \mathcal{Y}_{k},\mathcal{S}_{k}  \big)\;\;\;\;\;\;\;\;\;\;\;\;\;\;\;\;\;\;\;\;\;\;\;\;\;\;\;\;\;\;\;\;\;\;\;\;\;\;\;\;\;\;\;\;\;\;\;\;\;\;\;\;\;\;\;\;\;\\+\mathcal{P}{\big(\mathcal{Y}_k|\mathcal{X}_k} \big)   \nabla_{\mathcal{I} \big(   \mathcal{Y}_{k},\mathcal{S}_{k}  \big) }^{(\alpha)}  \mathcal{V} _{A,1}\Big(k;\mathcal{P}{\big(\mathcal{Y}|\mathcal{X}} \big) \Big) +\mathcal{W}^{(A)}_{hjb,1}
 \;\;\;\;\;\;\;\;\;\;\;\;\;\;\\ =\mathbb{E}_{\mathcal{I} \big(   \mathcal{Y}_{k},\mathcal{S}_{k}  \big)  } \left \{   \mathcal{V}  _{A,1}\Big(k;\mathcal{P}{\big(\mathcal{Y}|\mathcal{X}} \big) \Big) \right \};
\;\;\;\;\;\;\;\;\;\;\;\;\;\;\;\;\;\;\;\;\;\;\;\;\;\;\;\;\;\;\;\;\;\;\;\;\;
\end{split}
\end{equation*}
and
\item \textit{(ii) for Bob as}
\begin{equation*}
\begin{split}
\; HJB_{B,1}\triangleq\; \mathcal{D}_k^{(\alpha)}   \left \{ \mathcal{V}  _{B,1}\Big(k;\mathcal{P}{\big(\mathcal{Y}|\mathcal{X}} \big) \Big)\right \}
\;\;\;\;\;\;\;\;\;\;\;\;\;\;\;\;\;\;\;\;\;\;\;\;\;\;\;\; \;\;\;\;\;\\
+ \mathop{{\rm min}}\limits_{\mathcal{P}{\big(\mathcal{Y}|\mathcal{X}} \big)} {\rm \;  \mathcal{I} \big(   \hat{\mathcal{X}}_{k},\mathcal{S}_{k}  \big)} \;\;\;\;\;\;\;\;\;\;\;\;\;\;\;\;\;\;\;\;\;\;\;\;\;\;\;\;\;\;\;\;\;\;\;\;\;\;\;\;\;\;\;\;\;\;\;\;\;\;\;\;\;\;\;\;\;\\ +{\mathcal{P}{\big(\mathcal{Y}_k|\mathcal{X}_k} \big)   \nabla_{\mathcal{I} \big(   \hat{\mathcal{X}}_{k},\mathcal{S}_{k}  \big) }^{(\alpha)}  \mathcal{V} _{B,1}\Big(k;\mathcal{P}{\big(\mathcal{Y}|\mathcal{X}} \big) \Big) }+\mathcal{W}^{(B)}_{hjb,1} 
 \;\;\;\;\;\;\;\;\;\;\;\;\;\;\\=\mathbb{E}_{\mathcal{I} \big(   \hat{\mathcal{X}}_{k},\mathcal{S}_{k}  \big)  } \left \{   \mathcal{V}  _{B,1}\Big(k;\mathcal{P}{\big(\mathcal{Y}|\mathcal{X}} \big) \Big)\right \}.
\;\;\;\;\;\;\;\;\;\;\;\;\;\;\;\;\;\;\;\;\;\;\;\;\;\;\;\;\;\;\;\;\;\;\;\;\;
\end{split}
\end{equation*}
\end{itemize}

\item \textcolor{red}{\textbf{(\textit{iv})}} \textit{\textsc{FPK:} The FPK equation is written: }
\begin{itemize}
\item \textit{(i) for Alice as}
\begin{equation*}
\begin{split}
\; FPK_{A,1}\triangleq \; \mathcal{D}_k^{(\alpha)}   \left \{\mathcal{P}df_{A,1} \right \}= \nabla_{\mathcal{I} \big(   \mathcal{Y}_{k},\mathcal{S}_{k}  \big) }^{(\alpha)}  \;\;\;\;\;\;\;\;\;\;\;\;\;\;\; \\  \left \{\mathcal{V} _{A,1}\Big(k;\mathcal{P}{\big(\mathcal{Y}|\mathcal{X}} \big) \Big) \mathcal{P}df _{A,1}\right \}+\mathcal{W}^{(A)}_{fpk,1};
\end{split}
\end{equation*}
and
\item \textit{(ii) for Bob as }
\begin{equation*}
\begin{split}
\; FPK_{B,1}\triangleq \; \mathcal{D}_k^{(\alpha)}   \left \{\mathcal{P}df_{B,1} \right \}= \nabla_{\mathcal{I} \big(   \hat{\mathcal{X}}_{k},\mathcal{S}_{k}  \big) }^{(\alpha)}  \;\;\;\;\;\;\;\;\;\;\;\;\;\;\; \\  \left \{\mathcal{V} _{B,1}\Big(k;\mathcal{P}{\big(\mathcal{Y}|\mathcal{X}} \big) \Big) \mathcal{P}df_{B,1} \right \}+\mathcal{W}^{(B)}_{fpk,1}.
\end{split}
\end{equation*}
\end{itemize}
\end{itemize}
\end{proposition}

\textbf{\textsc{Proof:}} The proof is performed in terms of the following 3-step solution.

\textsc{Step 1:} Alice and Bob respectively follow $\mathcal{P}r \left \{   \mathcal{I} \big(   \mathcal{Y}_{k},\mathcal{S}_{k}  \big) < \partial^{(k)}_3   \right \} \ge \gamma_1$ and $\mathcal{P}r \left \{   \mathcal{I} \big(   \mathcal{Y}_{k},\mathcal{S}_{k}  \big) >\partial^{(k)}_4   \right \} \ge \gamma^{\prime}_1$ where $\gamma_1,\gamma^{\prime}_1$ are some thresholds. This initially means that Bob is indeed interested in an increment in $\mathcal{I} \big(   \hat{\mathcal{X}}_{k},\mathcal{S}_{k}  \big) $. Here, we need to prove that $\mathcal{I} \big(   \hat{\mathcal{X}}_{k},\mathcal{S}_{k}  \big) $ is a function of both $\mathcal{I} \big(   \mathcal{Y}_{k},\mathcal{S}_{k}  \big) $ and $\mathcal{P} \big(   \mathcal{Y}_{k}|\mathcal{X}_{k}  \big)$ as follows. We know 
\begin{equation*}
\begin{split}
\; \mathcal{I} \big(   \hat{\mathcal{X}}_{k},\mathcal{S}_{k}  \big) \triangleq \mathbb{E}_{\mathcal{S}_{k}}\left \{     \mathcal{P} \big(   \mathcal{S}_{k}| \hat{\mathcal{X}}_{k}  \big)  ||  \mathcal{P} \big(   \mathcal{S}_{k}  \big)   \right \},
\end{split}
\end{equation*}
where 
\begin{equation*}
\begin{split}
\; \mathcal{P} \big(   \mathcal{S}_{k}| \hat{\mathcal{X}}_{k}  \big)\triangleq\sum_x \mathcal{P} \big(   \mathcal{S}_{k}| {\mathcal{X}}_{k}  \big) \mathcal{P} \big(   \mathcal{X}_{k}| \hat{\mathcal{X}}_{k}  \big),
\end{split}
\end{equation*}
and we know\footnote{See e.g. \cite{2}.}
\begin{equation*}
\begin{split}
\; \mathcal{P} \big(   \mathcal{S}_{k}| {\mathcal{X}}_{k}  \big)\triangleq\frac{\mathcal{P} \big(   \mathcal{S}_{k} \big)}{\mathbb{Z} \Big (  \mathcal{P} \big(  {\mathcal{X}}_{k}  \big);\theta (\cdot) \Big)}  exp \Big(   \mathcal{P} \big(   \mathcal{Y}_{k}| {\mathcal{S}}_{k}  \big) || \mathcal{P} \big(   \mathcal{Y}_{k}| {{\mathcal{X}}}_{k}  \big) \Big),
\end{split}
\end{equation*}
where $\mathbb{Z} \big( \cdot; \theta(\cdot) \big)$ is a normalisation factor. This completely proves that $\mathcal{I} \big(   \hat{\mathcal{X}}_{k},\mathcal{S}_{k}  \big) $ is a function of $\mathcal{P} \big(   \mathcal{Y}_{k}|\mathcal{X}_{k}  \big)$ as well as $\mathcal{P} \big(   \mathcal{Y}_{k}|\mathcal{S}_{k}  \big) $, i.e., $\mathcal{I} \big(   \mathcal{Y}_{k},\mathcal{S}_{k}  \big) $. 

\textsc{Step 2:} The terms
\begin{equation*}
\begin{split}
\; \mathcal{P}r \left \{   \mathcal{I} \big(   \mathcal{Y}_{k},\mathcal{S}_{k}  \big) < \partial^{(k)}_3   \right \} \ge \gamma_1,
\end{split}
\end{equation*}
and 
\begin{equation*}
\begin{split}
\; \mathcal{P}r \left \{   \mathcal{I} \big(   \mathcal{Y}_{k},\mathcal{S}_{k}  \big) >\partial^{(k)}_4   \right \} \ge \gamma^{\prime}_1,
\end{split}
\end{equation*}
also show the convexity and concavity of respectively $\mathcal{I} \big(   \mathcal{Y}_{k},\mathcal{S}_{k}  \big)$ and $   \mathcal{I} \big(   \hat{\mathcal{X}}_{k},\mathcal{S}_{k}  \big) $ w.r.t. $\mathcal{P} \big(   \mathcal{Y}_{k}|\mathcal{X}_{k}  \big)$, and one can also see that the $+/-$ symbols in the control-laws are obvious. 

\textsc{Step 3:} Thanks to Remark 1, one can write the MFG equations\footnote{See e.g. \cite{6, 7} in order to understand how to formulate HJB and FPK.}.  

This completes the proof.$\; \; \; \blacksquare$

\subsection{Our second MFG: $ \mathcal{I} \big(   \mathcal{X}_{k},\mathcal{Y}_{k}  \big) $}

\begin{proposition} \label{P2}
 \textit{Our second MFG in relation to the term $ \mathcal{I} \big(   \mathcal{Y}_{k},\mathcal{S}_{k}  \big) $ is presented as follows.} 
\begin{itemize}
\item \textcolor{red}{\textbf{(\textit{i})}} \textit{\textsc{Control-Law}: According to Lemma 1, one can write the following control-laws: }
\begin{itemize}
\item \textit{(i) for Alice as}
\begin{equation*}
\begin{split}
\; \mathcal{D}_k^{(\alpha)}   \left \{ \mathcal{I} \big(   \mathcal{X}_{k},\mathcal{Y}_{k}  \right \}  =\;\;\;\;\;\;\;\;\;\;\;\;\;\;\;\;\;\;\;\;\;\;\;\;\;\;\;\;\;\;\;\;\;\;\;\;\;\;\;\;\;\;\;\;\;\;\;\; \\- \Phi_3 \Big (\mathcal{P} \big(   \mathcal{Y}_{k}|\mathcal{X}_{k}  \big)  ; \Phi^{\prime}_3 (k;\alpha);k \Big)+ \mathcal{W}^{(k)}_3;
\end{split}
\end{equation*}
and
\item \textit{(ii) for Bob as}
\begin{equation*}
\begin{split}
\; \mathcal{D}_k^{(\alpha)}   \left \{ \;  \mathcal{I} \big(   \mathcal{Y}_{k},\mathcal{X}_{k}  \big) \right \} \propto  \mathcal{D}_k^{(\alpha)}  \left \{ \;  \mathcal{I} \big(   \hat{\mathcal{X}}_{k},\mathcal{X}_{k}  \big) \right \}  =\;\;\;\;\;\;\;\;\;\;\; \\+ \Phi_4 \Big (\mathcal{P} \big(   \mathcal{Y}_{k}|\mathcal{X}_{k}  \big)  ; \Phi^{\prime}_4 (k;\alpha);k \Big) + \mathcal{W}^{(k)}_4,
\end{split}
\end{equation*}
where $\mathcal{W}^{(k)}_i,i \in \{ 3,4 \}$ is a time-varying random walk Winner process, and we assume $|\Phi^{(k)}_i |>0,i\in \{3,4\}$ without loss of generality\footnote{In order to preserve the type of the law’s conservation/non-conservation.}. 
\end{itemize}

\item \textcolor{red}{\textbf{(\textit{ii})}} \textit{\textsc{Value function:} The value function is written: }
\begin{itemize}
\item \textit{(i) for Alice as }
\begin{equation*}
\begin{split}
\;  \mathcal{V} _{A,2}\Big(k;\mathcal{P}{\big(\mathcal{Y}|\mathcal{X}} \big) \Big)\triangleq\mathop{{\rm min}}\limits_{\mathcal{P}{\big(\mathcal{Y}|\mathcal{X}} \big)}\mathbb{E} \left \{ \int_0^\mathcal{K}  \mathcal{I} \big(   \mathcal{Y}_{k},\mathcal{X}_{k}  \big) dk \right \};
\end{split}
\end{equation*}
and

\item \textit{(ii) for Bob as }
\begin{equation*}
\begin{split}
\; \mathcal{V} _{B,2}\Big(k;\mathcal{P}{\big(\mathcal{Y}|\mathcal{X}} \big) \Big)\triangleq\mathop{{\rm max}}\limits_{\mathcal{P}{\big(\mathcal{Y}|\mathcal{X}} \big)}\mathbb{E} \left \{ \int_0^\mathcal{K} \mathcal{I} \big(   \hat{\mathcal{X}}_{k},\mathcal{X}_{k}  \big)dk \right \}.
\end{split}
\end{equation*}
\end{itemize}

\item \textcolor{red}{\textbf{(\textit{iii})}} \textit{\textsc{HJB:} The HJB equation is written: }
\begin{itemize}
\item \textit{(i) for Alice as}
\begin{equation*}
\begin{split}
\; HJB_{A,2}\triangleq\; \mathcal{D}_k^{(\alpha)}   \left \{ \mathcal{V}  _{A,2}\Big(k;\mathcal{P}{\big(\mathcal{Y}|\mathcal{X}} \big) \Big)\right \}\;\;\;\;\;\;\;\;\;\;\;\;\;\;\;\;\;\;\;\;\;\;\;\;\;\;\;\; \;\;\;\;\;\\+ \mathop{{\rm min}}\limits_{\mathcal{P}{\big(\mathcal{Y}|\mathcal{X}} \big)} \;  \mathcal{I} \big(   \mathcal{Y}_{k},\mathcal{X}_{k}  \big) \;\;\;\;\;\;\;\;\;\;\;\;\;\;\;\;\;\;\;\;\;\;\;\;\;\;\;\;\;\;\;\;\;\;\;\;\;\;\;\;\;\;\;\;\;\;\;\;\;\;\;\;\;\;\;\;\;\\ \\+\mathcal{P}{\big(\mathcal{Y}_k|\mathcal{X}_k} \big)   \nabla_{\mathcal{I} \big(   \mathcal{Y}_{k},\mathcal{X}_{k}  \big) }^{(\alpha)}  \mathcal{V} _{A,2}\Big(k;\mathcal{P}{\big(\mathcal{Y}|\mathcal{X}} \big) \Big) +\mathcal{W}^{(A)}_{hjb,2}  \;\;\;\;\;\;\;\;\;\;\;\;\;\;\\=\mathbb{E}_{\mathcal{I} \big(   \mathcal{Y}_{k},\mathcal{X}_{k}  \big)  } \left \{   \mathcal{V}  _{A,2}\Big(k;\mathcal{P}{\big(\mathcal{Y}|\mathcal{X}} \big) \Big) \right \};\;\;\;\;\;\;\;\;\;\;\;\;\;\;\;\;\;\;\;\;\;\;\;\;\;\;\;\;\;\;\;\;\;\;\;\;\;
\end{split}
\end{equation*}
and

\textit{\item (ii) for Bob as}
\begin{equation*}
\begin{split}
\; HJB_{B,2}\triangleq\; \mathcal{D}_k^{(\alpha)}   \left \{ \mathcal{V}  _{B,2}\Big(k;\mathcal{P}{\big(\mathcal{Y}|\mathcal{X}} \big) \Big)\right \}\;\;\;\;\;\;\;\;\;\;\;\;\;\;\;\;\;\;\;\;\;\;\;\;\;\;\;\; \;\;\;\;\;\\+\mathop{{\rm min}}\limits_{\mathcal{P}{\big(\mathcal{Y}|\mathcal{X}} \big)}  \;  \mathcal{I} \big(   \hat{\mathcal{X}}_{k},\mathcal{X}_{k}  \big) \;\;\;\;\;\;\;\;\;\;\;\;\;\;\;\;\;\;\;\;\;\;\;\;\;\;\;\;\;\;\;\;\;\;\;\;\;\;\;\;\;\;\;\;\;\;\;\;\;\;\;\;\;\;\;\;\;\\ +\mathcal{P}{\big(\mathcal{Y}_k|\mathcal{X}_k} \big)   \nabla_{\mathcal{I} \big(   \hat{\mathcal{X}}_{k},\mathcal{X}_{k}  \big) }^{(\alpha)}  \mathcal{V} _{B,2}\Big(k;\mathcal{P}{\big(\mathcal{Y}|\mathcal{X}} \big) \Big) +\mathcal{W}^{(B)}_{hjb,2} \;\;\;\;\;\;\;\;\;\;\;\;\;\;\\=\mathbb{E}_{\mathcal{I} \big(   \hat{\mathcal{X}}_{k},\mathcal{X}_{k}  \big)  } \left \{   \mathcal{V}  _{B,2}\Big(k;\mathcal{P}{\big(\mathcal{Y}|\mathcal{X}} \big) \Big) \right \}.\;\;\;\;\;\;\;\;\;\;\;\;\;\;\;\;\;\;\;\;\;\;\;\;\;\;\;\;\;\;\;\;\;\;\;\;\;
\end{split}
\end{equation*}
\end{itemize}

\item \textcolor{red}{\textbf{(\textit{iv})}} \textit{\textsc{FPK:} The FPK equation is written: }
\begin{itemize}
\item \textit{(i) for Alice as }
\begin{equation*}
\begin{split}
\; FPK_{A,2}\triangleq\; \mathcal{D}_k^{(\alpha)}   \left \{\mathcal{P}df_{A,2} \right \}= \nabla_{\mathcal{I} \big(   \mathcal{X}_{k},\mathcal{Y}_{k}  \big) }^{(\alpha)} \;\;\;\;\;\;\;\;\;\;\;\;\;\;\;  \\  \left \{\mathcal{V} _{A,2}\Big(k;\mathcal{P}{\big(\mathcal{Y}|\mathcal{X}} \big) \Big) \mathcal{P}df_{A,2} \right \}+\mathcal{W}^{(A)}_{fpk,2};
\end{split}
\end{equation*}
and
\item \textit{(ii) for Bob as}
\begin{equation*}
\begin{split}
\; FPK_{B,2}\triangleq\; \mathcal{D}_k^{(\alpha)}   \left \{\mathcal{P}df_{B,2} \right \}= \nabla_{\mathcal{I} \big(   \hat{\mathcal{X}}_{k},\mathcal{X}_{k}  \big) }^{(\alpha)} \;\;\;\;\;\;\;\;\;\;\;\;\;\;\; \\ \left \{\mathcal{V} _{B,2}\Big(k;\mathcal{P}{\big(\mathcal{Y}|\mathcal{X}} \big) \Big) \mathcal{P}df_{B,2} \right \}+\mathcal{W}^{(B)}_{fpk,2}.
\end{split}
\end{equation*}

\end{itemize}
\end{itemize}
\end{proposition}

\textbf{\textsc{Proof:}} The proof is performed in terms of the following 3-step solution.

\textsc{Step 1:} Alice and Bob respectively follow $\mathcal{P}r \left \{   \mathcal{I} \big(   \mathcal{Y}_{k},\mathcal{X}_{k}  \big) < \partial^{(k)}_2   \right \} \ge \gamma_2$ and $\mathcal{P}r \left \{   \mathcal{I} \big(   \mathcal{Y}_{k},\mathcal{X}_{k}  \big) >\partial^{(k)}_5   \right \} \ge \gamma^{\prime}_2$ where $ \gamma_2, \gamma^{\prime}_2$ are some thresholds. This initially means that Bob is indeed interested in an increment in $\mathcal{I} \big(   \hat{\mathcal{X}}_{k},\mathcal{X}_{k}  \big) $. Here, we need to prove that $\mathcal{I} \big(   \hat{\mathcal{X}}_{k},\mathcal{X}_{k}  \big) $ is a function of $\mathcal{I} \big(   \mathcal{Y}_{k},\mathcal{X}_{k}  \big) $, i.e., $\mathcal{P} \big(   \mathcal{Y}_{k}|\mathcal{X}_{k}  \big)$ as follows. We know 
\begin{equation*}
\begin{split}
\; \mathcal{I} \big(   \hat{\mathcal{X}}_{k},\mathcal{X}_{k}  \big) \triangleq \mathbb{E}_{\mathcal{X}_{k}}\left \{     \mathcal{P} \big(   \mathcal{X}_{k}| \hat{\mathcal{X}}_{k}  \big)  ||  \mathcal{P} \big(   \mathcal{X}_{k}  \big)   \right \},
\end{split}
\end{equation*}
and we know 
\begin{equation*}
\begin{split}
\; \mathcal{P} \big(   \mathcal{X}_{k}| {\hat{\mathcal{X}}}_{k}  \big)\triangleq\frac{\mathcal{P} \big(   \mathcal{X}_{k} \big)}{\mathbb{Z} \Big (  \mathcal{P} \big(  \hat{{\mathcal{X}}}_{k}  \big);\theta (\cdot) \Big)}  exp \Big(   \mathcal{P} \big(   \mathcal{S}_{k}| {\mathcal{X}}_{k}  \big) || \mathcal{P} \big(   \mathcal{S}_{k}| \hat{{{\mathcal{X}}}}_{k}  \big) \Big),
\end{split}
\end{equation*}
where $\mathbb{Z} \big( \cdot; \theta(\cdot) \big)$ is a normalisation factor. The rest is totally similar to the proof of Proposition 1. 

\textsc{Step 2:} $\mathcal{P}r \left \{   \mathcal{I} \big(   \mathcal{Y}_{k},\mathcal{S}_{k}  \big) < \partial^{(k)}_3   \right \} \ge \gamma_1$ and $\mathcal{P}r \left \{   \mathcal{I} \big(   \mathcal{Y}_{k},\mathcal{S}_{k}  \big) >\partial^{(k)}_4   \right \} \ge \gamma^{\prime}_1$ also shows the concavity and convexity of respectively $\mathcal{I} \big(   \hat{\mathcal{X}}_{k},\mathcal{X}_{k}  \big)$ and $   \mathcal{I} \big(   \mathcal{Y}_{k},\mathcal{X}_{k}  \big) $ w.r.t. $\mathcal{P} \big(   \mathcal{Y}_{k}|\mathcal{X}_{k}  \big)$, and one can also see that the $+/-$ symbols in the control-laws are obvious. 

\textsc{Step 3:} Thanks to Remark 1, one can write the MFG equations.  

This completes the proof.$\; \; \; \blacksquare$

\subsection{Further discussions}
\begin{theorem} \label{T1}
 \textit{Our first MFG in relation to the term $ \mathcal{I} \big(   \mathcal{Y}_{k},\mathcal{S}_{k}  \big) $ can be jointly represented as follows.}
\begin{itemize}
\item \textcolor{red}{\textbf{(\textit{i})}} \textit{\textsc{Control-Law}: One can write the control-law as}
\begin{equation*}
\begin{split}
\; \mathcal{D}_k^{(\alpha)}   \left \{ \mathcal{I} \big(   \mathcal{Y}_{k},\mathcal{S}_{k}  \big)  \right \}  = - \Phi_1 \Big (\mathcal{P} \big(   \mathcal{Y}_{k}|\mathcal{X}_{k}  \big)  ; \Phi^{\prime}_1 (k;\alpha);k \Big)\\+\Phi_2 \Big (\mathcal{P} \big(   \mathcal{Y}_{k}|\mathcal{X}_{k}  \big)  ; \Phi^{\prime}_2 (k;\alpha);k \Big)\\+ \mathcal{W}^{(k)}_5.\;\;\;\;\;\;\;\;\;\;\;\;\;\;\;\;\;\;\;\;\;\;\;\;\;\;\;\;\;\;\;
\end{split}
\end{equation*}

\item \textcolor{red}{\textbf{(\textit{ii})}} \textit{\textsc{Value function:} The value function is written as }
\begin{equation*}
\begin{split}
\; \mathcal{V} _{1}\Big(k;\mathcal{P}{\big(\mathcal{Y}|\mathcal{X}} \big) \Big)\triangleq\;\;\;\;\;\; \;\;\;\;\;\; \;\;\;\;\;\; \;\;\;\;\;\; \;\;\;\;\;\; \;\;\;\;\;\; \;\;\;\;\;\; \;\;\;\;\;\; \;\; \;\;\;\;\;\ \\ \mathcal{V} _{A,1}\Big(k;\mathcal{P}{\big(\mathcal{Y}|\mathcal{X}} \big) \Big)+ \mathcal{V} _{B,1}\Big(k;\mathcal{P}{\big(\mathcal{Y}|\mathcal{X}} \big) \Big)=\mathop{{\rm min}}\limits_{\mathcal{P}{\big(\mathcal{Y}|\mathcal{X}} \big)} \cdots \;\;\;\;\;\;\; \\\mathbb{E} \left \{ \int_0^\mathcal{K}  \mathcal{I} \big(   \mathcal{Y}_{k},\mathcal{S}_{k}  \big) dk \right \}+\mathop{{\rm max}}\limits_{\mathcal{P}{\big(\mathcal{Y}|\mathcal{X}} \big)}\mathbb{E} \left \{ \int_0^\mathcal{K} \mathcal{I} \big(   \hat{\mathcal{X}}_{k},\mathcal{S}_{k}  \big)dk \right \}.
\end{split}
\end{equation*}

\item \textcolor{red}{\textbf{(\textit{iii})}} \textit{\textsc{HJB:} The HJB equation is written as}
\begin{equation*}
\begin{split}
\; HJB_{1}\triangleq\; \mathcal{D}_k^{(\alpha)}   \left \{ \mathcal{V}_{1} \Big(k;\mathcal{P}{\big(\mathcal{Y}|\mathcal{X}} \big) \Big)\right \}\;\;\;\;\;\;\;\;\;\;\;\;\;\;\;\;\;\;\;\;\;\;\;\;\;\;\;\;\;\;\;\;\\+ \mathop{{\rm min}}\limits_{\mathcal{P}{\big(\mathcal{Y}|\mathcal{X}} \big)}\mathop{{\rm max}}\limits_{\mathcal{P}{\big(\mathcal{S}|\hat{\mathcal{X}}} \big)}\mathcal{I} \big(   \mathcal{Y}_{k},\mathcal{S}_{k}  \big)  \cdots \;\;\;\;\;\;\; \;\;\;\;\;\;\;\;\;\;\;\;\;\;\;\;\;\;\;\;\;\;\;\;\;\;\\\rm \;  +\mathcal{P}{\big(\mathcal{Y}_k|\mathcal{X}_k} \big)   \nabla_{\mathcal{I} \big(   \mathcal{Y}_{k},\mathcal{S}_{k}  \big) }^{(\alpha)}  \nabla_{\mathcal{I} \big(   \hat{\mathcal{X}}_{k},\mathcal{S}_{k}  \big) }^{(\alpha)}     \mathcal{V} _{1}\Big(k;\mathcal{P}{\big(\mathcal{Y}|\mathcal{X}} \big) \Big)\;\;\;\;\;\;\;\;\\+\mathcal{W}_{hjb,1} =\mathbb{E}_{\mathcal{I} \big(   \mathcal{Y}_{k},\mathcal{S}_{k}  \big)  } \mathbb{E}_{\mathcal{I} \big(   \hat{\mathcal{X}}_{k},\mathcal{S}_{k}  \big)  } \left \{   \mathcal{V} _{1} \Big(k;\mathcal{P}{\big(\mathcal{Y}|\mathcal{X}} \big) \Big) \right \}.
\end{split}
\end{equation*}

\item \textcolor{red}{\textbf{(\textit{iv})}} \textit{\textsc{FPK:} The FPK equation is written as }
\begin{equation*}
\begin{split}
\; FPK_{1}\triangleq \; \mathcal{D}_k^{(\alpha)}   \left \{\mathcal{P}df_1 \right \}=\;\;\;\;\;\;\;\;\;\;\;\;\;\;\;\;\;\;\;\;\;\;\;\;\;\;\;\;\;\;\;\;\;\;\;\;\;\;\;\;\;\;\;\;\;\\ \nabla_{\mathcal{I} \big(   \mathcal{Y}_{k},\mathcal{S}_{k}  \big) }^{(\alpha)}   \nabla_{\mathcal{I} \big(   \hat{\mathcal{X}}_{k},\mathcal{S}_{k}  \big) }^{(\alpha)}  \left \{\mathcal{V} _{1}\Big(k;\mathcal{P}{\big(\mathcal{Y}|\mathcal{X}} \big) \Big) \mathcal{P}df _1\right \}+\mathcal{W}_{fpk,1}.
\end{split}
\end{equation*}
\end{itemize}
\end{theorem}

\textbf{\textsc{Proof:}} The proof is straightforward and easy to follow. The proof also follows the proof of Proposition 1. The term $\mathop{{\rm max}}\limits_{(\cdot)}\mathop{{\rm min}}\limits_{(\cdot)}$ additionally shows a Nash equilibrium as a no-lose-no-win or win-win or lose-lose game as the saddle-point. Meanwhile, the terms $\nabla_{(\cdot) }^{(\alpha)} \nabla_{(\cdot) }^{(\alpha)}$ and $\mathbb{E}_{(\cdot) }\mathbb{E}_{(\cdot) }$ $-$ or accordingly $\mathbb{E}_{(\cdot) ,(\cdot) }$ $-$ stand for the fact that the PDF of the MFG is now a 2-D joint one. 

In addition, as discussed, we have two control variables $\theta_{k,1}^{(\ell)}$ and $\theta_{k,2}^{(\ell)}$. So, one can consider the control-law $\dot{\mathscr{J}}=\mathscr{A}\mathscr{J}+\mathscr{B}_1 \theta_{k,1}^{(\ell)}+\mathscr{B}_2 \theta_{k,2}^{(\ell)}+\mathscr{W}_0$, consequently, formulating the value function
\begin{equation*}
\begin{split}
\; \mathop{{\rm max}}\limits_{ \theta_{k,1}^{(\ell)}  }\mathop{{\rm min}}\limits_{ \theta_{k,2}^{(\ell)}  } \left \{  \pi \big(  \theta_{k,1}^{(\ell)};\theta_{k,2}^{(\ell)} \big)=\pi \Big( \left | \theta_{k,1}^{(\ell)} \right |^2 -\left | \theta_{k,2}^{(\ell)} \right |^2 \Big)  \right \},
\end{split}
\end{equation*}
where 
\begin{equation*}
\begin{split}
\; \pi \big(  \theta_{k,1}^{\star(\ell)};\theta_{k,2}^{(\ell)} \big) \le  \pi \big(  \theta_{k,1}^{\star(\ell)};\theta_{k,2}^{\star(\ell)} \big) \le  \pi \big(  \theta_{k,1}^{(\ell)};\theta_{k,2}^{\star(\ell)} \big),
\end{split}
\end{equation*}
holds, where $(\cdot)^{\star}$ stands for the optimum value.

This completes the proof.$\; \; \; \blacksquare$

\begin{theorem} \label{T2}
 \textit{Our second MFG in relation to the term $ \mathcal{I} \big(   \mathcal{Y}_{k},\mathcal{S}_{k}  \big) $ can be jointly represented as follows.}
\begin{itemize}

\item \textcolor{red}{\textbf{(\textit{i})}} \textit{\textsc{Control-Law}: One can write the control-law as}
\begin{equation*}
\begin{split}
\; \mathcal{D}_k^{(\alpha)}   \left \{ \mathcal{I} \big(   \mathcal{X}_{k},\mathcal{Y}_{k}  \right \} = - \Phi_3 \Big (\mathcal{P} \big(   \mathcal{Y}_{k}|\mathcal{X}_{k}  \big)  ; \Phi^{\prime}_3 (k;\alpha);k \Big)+\\\Phi_4 \Big (\mathcal{P} \big(   \mathcal{Y}_{k}|\mathcal{X}_{k}  \big)  ; \Phi^{\prime}_4 (k;\alpha);k \Big)+ \mathcal{W}^{(k)}_6.
\end{split}
\end{equation*}

\begin{algorithm*}
\caption{\textcolor{black}{Federated learning algorithm to find an optimal solution to the graph $\mathscr{G}_m (\mathscr{Q},\mathscr{E}_m)$. }}\label{ffff}
\begin{algorithmic}
\STATE \textbf{\textsc{Initialisation.}} 

$\;\;$ \textbf{while} $\mathbb{TRUE}$ \textbf{do}

$     \;\;\;  \;  \;  $ (\textit{i}) Send Local observation path tuple $\Upsilon_2$ to the Header;

$     \;  \;  \;\;\;  $ (\textit{ii}) Alice and Bob respectively check if $\mathcal{I} \big(   \mathcal{X}^{(\ell)}_{k},\mathcal{S}^{(\ell^{\prime})}_{k^{\prime} } \big) \neq 0$, and $ \mathcal{I} \big(   \mathcal{Y}^{(\ell)}_{k},\mathcal{S}^{(\ell^{\prime} )}_{k^{\prime}}  \big) \neq 0$;

$     \;  \;  \;  \;       \;  $ and (\textit{iii}) Alice and Bob jointly apply Theorems 1 and 2.

$\;\;$ \textbf{endwhile} 

\textbf{end}

\textbf{\textsc{Output:}} $\mathcal{P}\big( \mathcal{Y}|\mathcal{X} \big)$
\end{algorithmic}
\end{algorithm*}

\item \textcolor{red}{\textbf{(\textit{ii})}} \textit{\textsc{Value function:} The value function is written as} 
\begin{equation*}
\begin{split}
\; \mathcal{V} _{2}\Big(k;\mathcal{P}{\big(\mathcal{Y}|\mathcal{X}} \big) \Big)\triangleq \;\;\;\;\;\;\;\;\;\;\;\;\;\;\;\;\;\;\;\;\;\;\;\;\;\;\;\;\;\;\;\;\;\;\;\;\;\;\;\;\;\;\;\;\;\;\;\;\;\;\;\;\;\;\;\\ \mathcal{V} _{A,2}\Big(k;\mathcal{P}{\big(\mathcal{Y}|\mathcal{X}} \big) \Big)+\mathcal{V} _{B,2}\Big(k;\mathcal{P}{\big(\mathcal{Y}|\mathcal{X}} \big) \Big)\\=\mathop{{\rm min}}\limits_{\mathcal{P}{\big(\mathcal{Y}|\mathcal{X}} \big)}\mathbb{E} \left \{ \int_0^\mathcal{K}  \mathcal{I} \big(   \mathcal{Y}_{k},\mathcal{X}_{k}  \big) dk \right \} \;\;\;\;\;\;\;\;\;\\ +\mathop{{\rm max}}\limits_{\mathcal{P}{\big(\hat{\mathcal{X}}|\mathcal{X}} \big)}\mathbb{E} \left \{ \int_0^\mathcal{K} \mathcal{I} \big(   \hat{\mathcal{X}}_{k},\mathcal{X}_{k}  \big)dk \right \}.\;\;\;\;\;\;\;\;
\end{split}
\end{equation*}

\item \textcolor{red}{\textbf{(\textit{iii})}} \textit{\textsc{HJB:} The HJB equation is written as} 
\begin{equation*}
\begin{split}
\; HJB_{2}\triangleq\; \mathcal{D}_k^{(\alpha)}   \left \{ \mathcal{V} _{2}  \Big(k;\mathcal{P}{\big(\mathcal{Y}|\mathcal{X}} \big) \Big)\right \}+ \mathop{{\rm min}}\limits_{\mathcal{P}{\big(\mathcal{Y}|\mathcal{X}} \big)} \mathop{{\rm max}}\limits_{\mathcal{P}{\big(\hat{\mathcal{X}}|\mathcal{X}} \big)} \\ \rm \;  \mathcal{I} \big(   \hat{\mathcal{X}}_{k},\mathcal{X}_{k}  \big) +\mathcal{P}{\big(\mathcal{Y}_k|\mathcal{X}_k} \big)   \nabla_{\mathcal{I} \big(   \mathcal{Y}_{k},\mathcal{X}_{k}  \big) }^{(\alpha)} \nabla_{\mathcal{I} \big(   \hat{\mathcal{X}}_{k},\mathcal{X}_{k}  \big) }^{(\alpha)} \\ \mathcal{V}  _{2}\Big(k;\mathcal{P}{\big(\mathcal{Y}|\mathcal{X}} \big) \Big)+\mathcal{W}_{hjb,2} \;\;\;\;\;\;\\=\mathbb{E}_{\mathcal{I} \big(   \hat{\mathcal{X}}_{k},\mathcal{X}_{k}  \big)  }\mathbb{E}_{\mathcal{I} \big(   \mathcal{Y}_{k},\mathcal{X}_{k}  \big)  } \left \{   \mathcal{V} _{2}  \Big(k;\mathcal{P}{\big(\mathcal{Y}|\mathcal{X}} \big) \Big)\right \}.\;\;\;
\end{split}
\end{equation*}


\item \textcolor{red}{\textbf{(\textit{iv})}} \textit{\textsc{FPK:} The FPK equation is written as }
\begin{equation*}
\begin{split}
\;FPK_{2}\triangleq\; \mathcal{D}_k^{(\alpha)}   \left \{\mathcal{P}df_2 \right \}= \nabla_{\mathcal{I} \big(   \hat{\mathcal{X}}_{k},\mathcal{X}_{k}  \big) }^{(\alpha)}  \nabla_{\mathcal{I} \big(   \mathcal{X}_{k},\mathcal{Y}_{k}  \big) }^{(\alpha)}   \\ \left \{\mathcal{V} _{2}\Big(k;\mathcal{P}{\big(\mathcal{Y}|\mathcal{X}} \big) \Big) \mathcal{P}df_2 \right \}+\mathcal{W}_{fpk,2}.\;\;\;
\end{split}
\end{equation*}
\end{itemize}
\end{theorem}

\textbf{\textsc{Proof:}} The proof is straightforward and totally similar to Theorem 1.$\; \; \; \blacksquare$

\begin{theorem} \label{T3}
 \textit{In relation to our joint MFGs for both first and second ones, $\mathcal{V} _{1}\Big(k;\mathcal{P}{\big(\mathcal{Y}|\mathcal{X}} \big) \Big) \neq 0$, $\mathcal{V} _{2}\Big(k;\mathcal{P}{\big(\mathcal{Y}|\mathcal{X}} \big) \Big) \neq0$.}
\end{theorem}

\textbf{\textsc{Proof:}} For $\mathcal{V} _{A,1}\Big(k;\mathcal{P}{\big(\mathcal{Y}|\mathcal{X}} \big) \Big)+\mathcal{V} _{B,1}\Big(k;\mathcal{P}{\big(\mathcal{Y}|\mathcal{X}} \big) \Big)$ we should examine $\mathbb{E} \left \{ \int_0^\mathcal{K} \mathcal{I} \big( \mathcal{Y}_{k},\mathcal{S}_{k} \big) dk \right \} - \mathbb{E} \left \{ \int_0^\mathcal{K} \mathcal{I} \big( \hat{\mathcal{X}}_{k},\mathcal{S}_{k} \big)dk \right \}$. The sign $-$ is applied due to the fact that one term is minimised, inversely, one should be maximised, that is, due to convexity of one term in addition to the concavity of another one, as discussed before. It should be noted that for this, we need $log \frac{\mathcal{P} \big( \mathcal{S}_{k}|\mathcal{Y}_{k} \big) }{\mathcal{P} \big( \mathcal{S}_{k}|\hat{\mathcal{X}}_{k} \big)}$. For the nominator, we need $ \left \{ \mathcal{P} \big( \mathcal{Y}_{k}|\mathcal{S}_{k} \big) || \mathcal{P} \big( \mathcal{Y}_{k}|\mathcal{Y}_{k} \big)  \right \}$ and the denominator we need $ \left \{  \mathcal{P} \big( \mathcal{Y}_{k}|\mathcal{S}_{k} \big) || \mathcal{P} \big( \mathcal{Y}_{k}|\hat{\mathcal{X}}_{k} \big) \right \}$. It is then entailed to have $\mathcal{P} \big( \mathcal{Y}_{k}|\hat{\mathcal{X}}_{k} \big)=1$ in the sense that $\mathcal{V} _{1}\Big(k;\mathcal{P}{\big(\mathcal{Y}|\mathcal{X}} \big) \Big)=0$ can be satisfied, something that is not going to happen. This means that the game is not zero-sum one as discussed in Lemma 1.

 For $\mathcal{V} _{A,2}\Big(k;\mathcal{P}{\big(\mathcal{Y}|\mathcal{X}} \big) \Big)+\mathcal{V} _{B,2}\Big(k;\mathcal{P}{\big(\mathcal{Y}|\mathcal{X}} \big) \Big)$ we should examine $\mathbb{E} \left \{ \int_0^\mathcal{K} \mathcal{I} \big( \mathcal{Y}_{k},\mathcal{X}_{k} \big) dk \right \}-\mathbb{E} \left \{ \int_0^\mathcal{K} \mathcal{I} \big( \hat{\mathcal{X}}_{k},\mathcal{X}_{k} \big)dk \right \}$. The sign $-$ is applied due to the fact that one term is minimised, inversely, one should be maximised, that is, due to convexity of one term in addition to the concavity of another one, as discussed above. It should be noted that for this, we need $log \frac{\mathcal{P} \big( \mathcal{X}_{k}|\hat{\mathcal{X}_{k}} \big) }{\mathcal{P} \big( \mathcal{X}_{k}|\mathcal{Y}_{k} \big)}$. For the nominator, we need $ \left \{ \mathcal{P} \big( \mathcal{Y}_{k}|\mathcal{X}_{k} \big) || \mathcal{P} \big( \mathcal{Y}_{k}|\hat{\mathcal{X}}_{k} \big)  \right \}$ and the denominator we need $ \left \{  \mathcal{P} \big( \mathcal{Y}_{k}|\mathcal{X}_{k} \big) || \mathcal{P} \big( \mathcal{Y}_{k}|\mathcal{Y}_{k} \big) \right \}$. It is then entailed to have $\mathcal{P} \big( \mathcal{Y}_{k}|\hat{\mathcal{X}}_{k} \big)=1$ in the sense that $\mathcal{V} _{2}\Big(k;\mathcal{P}{\big(\mathcal{Y}|\mathcal{X}} \big) \Big)=0$ can be satisfied, something that is not going to happen as discussed. This means that the game is not zero-sum one as discussed in Lemma 2.

The proof is completed.$\; \; \; \blacksquare$

\begin{figure}[t]
\centering
\subfloat[CDF of the iterations: for the convergence.]{\includegraphics[trim={{20mm} {68 mm} {23mm} {73mm}},clip,scale=0.43]{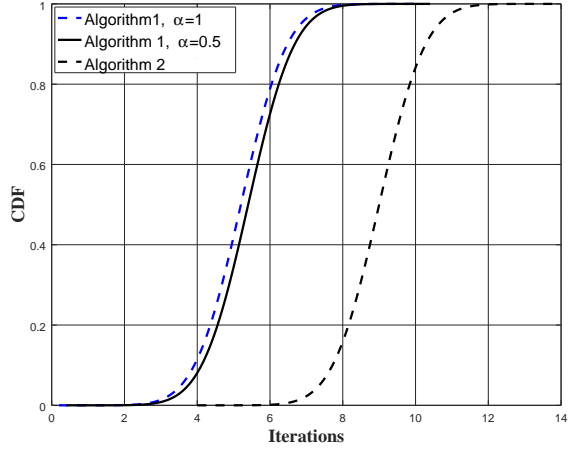}}\\
\subfloat[$\mathcal{I}\big(   \mathcal{S},\mathcal{Y} \big)$ v.s. the normalised version of $\mathcal{I}\big(   \mathcal{X},\mathcal{Y} \big)$.]{\includegraphics[trim={{14mm} {68 mm} {20mm} {73mm}},clip,scale=0.43]{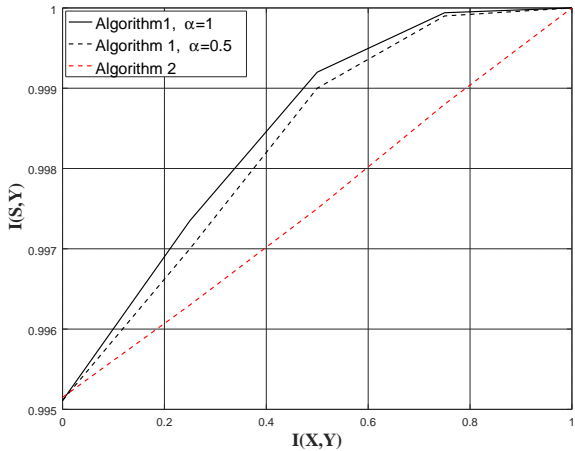}}\\
\subfloat[Loss (\%) v.s. the normalised iteration.]{\includegraphics[trim={{20mm} {68 mm} {23mm} {73mm}},clip,scale=0.43]{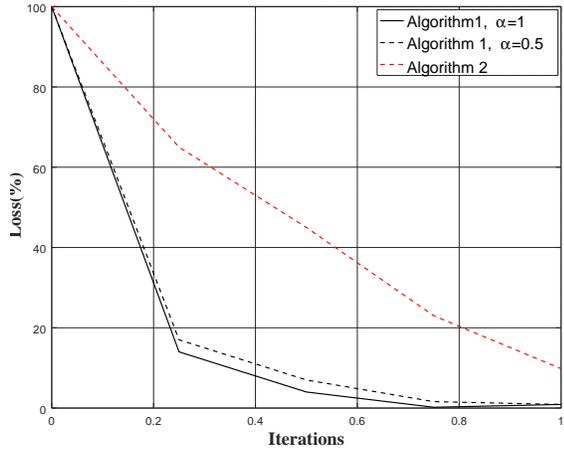}}
\caption{ Algorithms \ref{ffff} and \ref{f} while changing $\alpha$.} \label{F3}
\label{fig:EcUND} 
\end{figure} 

\begin{lemma} \label{L4}
\textit{ Our MFGs experience a Nash-equilibrium for themselves.}
\end{lemma}

\textbf{\textsc{Proof:}} The proof is straightforward and easy to follow. According to Theorem 3, one can see there is at least a no-lose-no-win or win-win or lose-lose case as the saddle-point which is satisfied between Alice and Bob. The proof is completed.$\; \; \; \blacksquare$

\begin{lemma} \label{L5}
\textit{ Our MFGs experience their steady-states since the system in both cases is stable.}
\end{lemma}

\textbf{\textsc{Proof:}} The proof is presented in the following.

\textit{First method:} According to Lemma 3 and regarding the fact that the system is in the sense that the hostility probability of Bob against Alice is neither $1$ nor $0$, so it is revealed that there exists a Nash-equilibrium in which the system can experience its steady-state.

\textit{Second method:} We know $-$ see e.g. \cite{16} $-$ one can formulate the stability criterion as follows $\int_0^{\infty} e^{-\rho t}\left |     \mathscr{F}(t+1)-  \mathscr{F}(t)  \right |^2 < \infty$. So, one can prove, according to the definition of Laplace transform $\mathscr{L}\{\cdot\}$, that it is equivalent to
\begin{equation*}
\begin{split}
\; \lim_{t \longrightarrow \infty}  \mathscr{F}(t) =\lim_{s \longrightarrow 0}  s \mathscr{L} \left \{ \mathscr{F}(t)  \right \} -\mathscr{F}(0^-)< \infty.
\end{split}
\end{equation*}
Moreover, we know $-$ see e.g. \cite{17} $-$ that $\mathcal{I}\big(\mathscr{f}, \mathscr{g}\big)$ or/and $\mathscr{f} || \mathscr{g}$ can be considered as the \textit{Radon-Nikodym} derivative $\frac{d \;\mathscr{f}}{d \;\mathscr{g}}$, so, one can reformulate the stability\footnote{Initially speaking, the three following cases for control over noisy communication channels generally exist \cite{18}: (\textit{i}) unstabilisable; (\textit{ii}) not completely controllable but stabilisable; and (\textit{iii}) controllable. Therefore, controllability, observability and detectability are defined as follows \cite{18}: \begin{itemize}\item A system $f$ is controllable if for any pair $(x_0,x_1) \in \mathbb{R}_n$ there exists $T>0$ and an input function $\theta(t) \in \mathbb{R}_m$ that
transfers the state $x(t)$ from $x_0$ at time $t=0$ to $x_1$ at time $t=T$. \item Observability means that for any input function, the current state can be determined in finite time using only the outputs. In fact, the system is observable if the output w.r.t. the $x_0$ is equal to the output w.r.t. $x_t, \forall t>0$ and
unobservable otherwise. \item A system is indeed detectable if we experience the output equating with zero for $x_0=0$, so, its steady state goes to zero, undetectable otherwise.   \end{itemize}} criterion as 
\begin{equation*}
\begin{split}
\mathscr{L} \left \{    \frac{\frac{d \;\mathscr{f}(t)}{dt}}{\frac{d \;\mathscr{g}(t)}{dt}}  =\frac{d \;\mathscr{f}(t)}{dt} \cdot \mathscr{h}(t)   \right \}=\frac{1}{2 \pi i}\int_{c-\infty}^{c+\infty} \cdots \;\;\;\; \;\;\; \;\;\; \\ \left \{\left \{ \lim_{s \longrightarrow 0} s \mathscr{L} \left \{ \mathscr{f}(t)  \right \} -\underbrace{\mathscr{f}(0^-)}_{0} \right \} \mathscr{L} \left \{   \mathscr{h} \big( \mathscr{g}(t);t\big)  \right \}\right \} < \infty,
\end{split}
\end{equation*}
which holds for our scheme since the hostility probability of Bob against Alice is neither $1$ nor $0$.

The proof is completed.$\; \; \; \blacksquare$

\begin{proposition} \label{P3}
\textit{One can consider the directed graph $\mathscr{G}_m (\mathcal{Q},\mathcal{E}_m)$ to which the optimal solution is guaranteed by Algorithm \ref{ffff}.}
\end{proposition} 

\textbf{\textsc{Proof:}} Recall Lemma 1 and its proof when we defined two groups of the major players, i.e., the Headers: (\textit{i}) at Alice $\mathcal{I} \big(   \mathcal{X}^{(\ell^{\prime})}_{k^{\prime}},\mathcal{S}^{(\ell^{\prime\prime} )}_{k^{\prime\prime} } \big) \neq 0$; and (\textit{ii}) at Bob $ \mathcal{I} \big(   \mathcal{Y}^{(\ell^{\prime})}_{k^{\prime}},\mathcal{S}^{(\ell^{\prime\prime} )}_{k^{\prime\prime}}  \big) \neq 0$. We denote $q^{(i)}_{\ell,k} \in \mathscr{Q}, \forall i \in \{A,B\}$ as the state of the agent $\ell$ in relation to either Alice, i.e., $i=A$ or Bob, i.e., $i=B$, at the time instant $k$. Indeed, any movement for the agents of Alice and Bob can be respectively the pairs $\big(   \mathcal{X}^{(\ell^{\prime})}_{k^{\prime}},\mathcal{S}^{(\ell^{\prime\prime} )}_{k^{\prime\prime} } \big)$ and $\big(   \mathcal{Y}^{(\ell^{\prime})}_{k^{\prime}},\mathcal{S}^{(\ell^{\prime\prime} )}_{k^{\prime\prime}}  \big)$. The movement of each agent is constrained by a directed graph $\mathscr{G}_m (\mathscr{Q},\mathscr{E}_m)$ where $\mathscr{E}_m \subseteq \mathscr{Q} \times \mathscr{Q}$. An agent can move from $q$ to $q^{\prime}$ in one time step if and only if the edge $\big(q,q^{\prime} \big) \in \mathscr{E}_m$. Obviously, $\big(q^{(i)}_{\ell,k},q^{i,\prime} _{\ell,k}\big) \in \mathscr{E}_m$ holds. The local state path is also denoted by the tuple 
\begin{equation*}
\begin{split}
\; \Upsilon_1\triangleq\bigg( \Big(q^{(i)}_{\ell,0} ,q^{(i)}_{\ell,1} ,\cdots \Big),\Big(q^{(i)}_{0,k} ,q^{(i)}_{1,k} ,\cdots \Big) \bigg),
\end{split}
\end{equation*}
and the local observation path is denoted by the tuple 
\begin{equation*}
\begin{split}
\; \Upsilon_2\triangleq\bigg( \Big(s^{(i)}_{\ell,0} ,s^{(i)}_{\ell,1} ,\cdots \Big),\Big(s^{(i)}_{0,k} ,s^{(i)}_{1,k} ,\cdots \Big) \bigg),
\end{split}
\end{equation*}
and the set of local state-observations is then defined by $\Omega_{\ell}\triangleq \left \{ \xi_{\ell}   |\xi_{\ell} =  \big(   \Upsilon_1,\Upsilon_2 \big) \right \} $. The global state observation paths is defined as $\Omega=	\prod_{\ell}\Omega_{\ell}$. 

The proof is then completed.$\; \; \; \blacksquare$

\subsection{Interpretation}
\subsubsection{A laminar flow of the information vs. a turbulent flow of the information}

\begin{proposition} \label{P3} \textit{The game discussed in this paper can be theoretically interpreted in the context of: (i) a laminar flow of the information; as well as (ii) a turbulent flow of the information.}\end{proposition} \label{P3}

\textsc{\textbf{Proof.}} Let us do an analysis from a \textit{turbulent flow} based point of view. The behaviour of a turbulent flow is in contrast to the laminar flow $-$ where the fluid smoothly undergoes in parallel layers. In fact, the turbulent flow (chaotic) is a type of fluid flow while the fluid physically experiences irregular fluctuations. 

Now, let us go over our pursuit-evasion game. In our mixed cooperative-and-non-cooperative game in relation to the two groups of techniques exemplified before, the story is that although Alice tries to send laminar flow to Bob, she is careful about Bob’s potentially adversarial issues against herself, thus, she tries to send in some points a turbulent flow to Bob. This kind of behaviour unhesitatingly entails a trade-off, which was gone over in-depth in the previous parts.

The proof is then completed.$\; \; \; \blacksquare$

\subsubsection{Incomplete information, Hyper-games, and Fuzzy competitive learning}

\begin{proposition} \label{P3} \textit{Incomplete information is theoretically analysable from another viewpoint, that is, a Hyper-game\footnote{Hyper-games model the interactions where one or more agents may follow totally different strategies. This is originated from the mis-perception of other players’ objectives and strategies. Hyper-games are information asymmetric based which are a branch of games with incomplete information where at least one agent gets access to the perfect information. Indeed, in a hyper-game scenario set, while they are free of taking into account a probabilistic uncertainty, agents are permitted to simultaneously report differences in their perceptions of the given game. In other words, a hyper-game is an extension of the concept of rationality to a bounded information sub-set \cite{19, 20, 21}.}-theoretical one in the context of Fuzzy competitive learning\footnote{The competitive learning issue is the one which investigates how well a specific learner (agent) is able to win. In the competitive learning paradigm, we cluster the training patterns into representative groups $-$ in the sense that patterns during a cluster, compared to the patterns belonging to totally different clusters, have the most similarities. Meanwhile, each training pattern belongs to a certain degree of the certain cluster in the fuzzy competitive learning paradigms, depending chiefly upon its distance to the center vector.} $-$ something that is of an absolutely relaxed nature.}\end{proposition} \label{P3}

\textsc{\textbf{Proof.}} The proof is given in the context of the following multi-step solution.

\textbf{\textsc{Definition 2. \textit{Hyper}-game (\cite{19, 20, 21})}} \textit{A level-1 two-player hyper-game is a pair $\mathscr{HG}^{(1)}=\langle \mathscr{G}_1, \mathscr{G}_2\rangle$, where $\mathscr{G}_1$, $\mathscr{G}_2$ are the games observed by the first and the second agents, respectively. A level-2 two-player hyper-game is a pair $\mathscr{HG}^{(2)} = \langle \mathscr{HG}^{(1)}, \mathscr{G}_2 \rangle$, where the first player observes the interaction as a level-1 hyper-game and the second player observes the interaction as the game $\mathscr{G}_2$.}

\textbf{\textsc{Remark 4 \cite{19, 20, 21}.}} \textit{Since the first agent is totally aware of $\mathscr{HG}^{(1)}$ whereas the second agent only knows $\mathscr{G}_2$, a level-2 hyper-game is absolutely sufficient to model the game with asymmetric information. Thus\footnote{It is totally obvious that the perception of the two players about the overall game is completely different.}, the first agent calculates the strategy by solving the hyper-game $\mathscr{HG}^{(1)}$ and the second player does the calculation by solving the game $\mathscr{G}_2$.}


\textsc{Step 1.} Let us call $\mathscr{S}^{(\varphi)}=\{ \mathscr{s}^{(\varphi)}_1; \cdots;\mathscr{s}^{(\varphi)}_k \} $ as a data set containing $k$ data points w.r.t. the given parameter $\varphi$, $q$ as the number of clusters which satisfies $2 \le q < k$, as well as $\mathcal{C}^{(\varphi)}_{\mathscr{F}}=\{ c^{(\varphi)}_1; \cdots; c^{(\varphi)}_q \} $ as a set of cluster centres, $\mu^{m,(\varphi)}_{ij} \in [0, 1]$ as the membership\footnote{See the prioneer work \cite{22} by the Father of Fuzzy mathematics, \textit{Zadeh}.} of $\mathscr{s}^{(\varphi)}_j$ in class $i$,
$m \in [1,\infty)$ is the degree of fuzzification\footnote{See the prioneer work \cite{22} by the Father of Fuzzy mathematics, \textit{Zadeh}.}, as well as the fact that $\mathscr{U} ^{(\varphi)}=\{ \mu^{m,(\varphi)}_{ij} \} \in \mathbb{R}^{ q \times k}$ is the membership matrix. It should be noted that the two given constraints strongly guarantee that each data has the same overall weight in the data set $-$ something which indicates that none of the clusters is supposed to be an empty set. 

\textsc{Step 2.} Let us sectionalise Bob into Bob and Eve here, due to the \textit{pursuit-evasion} nature expressed before.

Let us define the outage probability $\gamma$ for the Alice-Bob link. Let $\ell^{\star}$ be the solution for
\begin{equation*}
\begin{split}
\; \mathop{{\rm \mathbb{M}in}}\limits_{ \mathscr{U}^{(\varphi)}, \mathcal{C}^{(\varphi)}_{\mathscr{F}} } {\rm \; }\gamma \mathscr{J} \big( \mathscr{U}^{(\varphi)}, \mathcal{C}^{(\varphi)}_{\mathscr{F}} \big)+(1-\gamma)\mathscr{F}_1(\cdot),
\end{split}
\end{equation*}
for a given convex set $\mathscr{F}_1(\cdot)$. Inversely, Eve aims at, for a given concave set $\mathscr{F}_2(\cdot)$, 
\begin{equation*}
\begin{split}
\; \mathop{{\rm \mathbb{M}ax}}\limits_{ \mathscr{U}^{(\varphi)}, \mathcal{C}^{(\varphi)}_{\mathscr{F}} } {\rm \; } \gamma\mathscr{J} \big( \mathscr{U}^{(\varphi)}, \mathcal{C}^{(\varphi)}_{\mathscr{F}} \big)+(1-\gamma)\mathscr{F}_2(\cdot),
\end{split}
\end{equation*}
for which the solution is $\bar{\ell}^{\star}$.

 It should only be noted here that there exists a Stackelberg equilibrium where
\begin{equation*}
\begin{split}
\; \mathscr{J} ^{(Alice-Bob)}\big( \ell^{\star}; \mathscr{U}^{(\varphi)}, \mathcal{C}^{(\varphi)}_{\mathscr{F}} \big) \le \mathscr{J} \big( \ell;\bar{\ell}; \mathscr{U}^{(\varphi)}, \mathcal{C}^{(\varphi)}_{\mathscr{F}} \big),
\end{split}
\end{equation*}
and 
\begin{equation*}
\begin{split}
\; \mathscr{J} ^{(Eve)}\big( \bar{\ell}^{\star}; \mathscr{U}^{(\varphi)}, \mathcal{C}^{(\varphi)}_{\mathscr{F}} \big) \ge \mathscr{J} \big( \ell;\bar{\ell}; \mathscr{U}^{(\varphi)}, \mathcal{C}^{(\varphi)}_{\mathscr{F}} \big),
\end{split}
\end{equation*}
simultaneously hold. 

Let us define the outage probability $\gamma$ for the Alice-Bob link. Let $\ell^{\star}$ be the solution for
\begin{equation*}
\begin{split}
\; \mathop{{\rm \mathbb{M}in}}\limits_{ \mathscr{U}^{(\varphi)}, \mathcal{C}^{(\varphi)}_{\mathscr{F}} } {\rm \; }\gamma \mathscr{J} \big( \mathscr{U}^{(\varphi)}, \mathcal{C}^{(\varphi)}_{\mathscr{F}} \big)+(1-\gamma)\mathscr{F}_1(\cdot),
\end{split}
\end{equation*}
for a given convex set $\mathscr{F}_1(\cdot)$. Inversely, Eve aims at, for a given concave set $\mathscr{F}_2(\cdot)$, 
\begin{equation*}
\begin{split}
\; \mathop{{\rm \mathbb{M}ax}}\limits_{ \mathscr{U}^{(\varphi)}, \mathcal{C}^{(\varphi)}_{\mathscr{F}} } {\rm \; } \gamma\mathscr{J} \big( \mathscr{U}^{(\varphi)}, \mathcal{C}^{(\varphi)}_{\mathscr{F}} \big)+(1-\gamma)\mathscr{F}_2(\cdot),
\end{split}
\end{equation*}
for which the solution is $\bar{\ell}^{\star}$.

 It should only be noted here that there exists a Stackelberg equilibrium where
\begin{equation*}
\begin{split}
\; \mathscr{J} ^{(Alice-Bob)}\big( \ell^{\star}; \mathscr{U}^{(\varphi)}, \mathcal{C}^{(\varphi)}_{\mathscr{F}} \big) \le \mathscr{J} \big( \ell;\bar{\ell}; \mathscr{U}^{(\varphi)}, \mathcal{C}^{(\varphi)}_{\mathscr{F}} \big),
\end{split}
\end{equation*}
and 
\begin{equation*}
\begin{split}
\; \mathscr{J} ^{(Eve)}\big( \bar{\ell}^{\star}; \mathscr{U}^{(\varphi)}, \mathcal{C}^{(\varphi)}_{\mathscr{F}} \big) \ge \mathscr{J} \big( \ell;\bar{\ell}; \mathscr{U}^{(\varphi)}, \mathcal{C}^{(\varphi)}_{\mathscr{F}} \big),
\end{split}
\end{equation*}
simultaneously hold. 

\textsc{Step 3.} Take into account a set of players 
\begin{equation*}
\begin{split}
\; \mathscr{P}_1= \{\mathscr{P}_1(k);\mathscr{P}_1(\theta);\mathscr{P}_1(\ell);\cdot    \},
\end{split}
\end{equation*}
relating to the Alice-Bob link, as well as
\begin{equation*}
\begin{split}
\; \mathscr{P}_2= \{\mathscr{P}_2(k);\mathscr{P}_2(\theta);\mathscr{P}_2(\ell);\cdot    \},
\end{split}
\end{equation*}
relating to Eve’s one. Recall $\mathscr{F}_1$ and $\mathscr{F}_2$. For the arbitrary positive scalars $\gamma_i, i\in \{  1, 2, 3 \}$, $\mathscr{P}_1$ follows a competitive learning of 
\begin{equation*}
\begin{split}
\; \mathop{{\rm \mathbb{M}in}}\limits_{ \mathscr{U}^{(\varphi)}, \mathcal{C}^{(\varphi)}_{\mathscr{F}} } {\rm \; } \gamma_1\mathscr{J} \big( \mathscr{U}^{(\varphi)}, \mathcal{C}^{(\varphi)}_{\mathscr{F}} \big)+\gamma_2\mathscr{F}_1(\cdot)-\gamma_3\mathscr{F}_2(\cdot).
\end{split}
\end{equation*}
Regarding the fact that $-\gamma_3\mathscr{F}_2(\cdot)$ is a convex set and an empty set is the least convex set, we can re-write the later expression as the Problem $\mathscr{P}_2$ as
\begin{equation*}
\begin{split}
\; \big (\mathscr{P}_2 \big):\mathop{{\rm \mathbb{M}in}}\limits_{ \mathscr{U}^{(\varphi)}, \mathcal{C}^{(\varphi)}_{\mathscr{F}} } {\rm \; } \alpha_1\mathscr{J} \big( \mathscr{U}^{(\varphi)}, \mathcal{C}^{(\varphi)}_{\mathscr{F}} \big)+\alpha_2\mathscr{F}_1(\cdot),\;\;\;\;\;\;\;\;\;\;\;\;\;\;\;\;\;\;\;\;\;\; \\
s.t. \;\; \mathscr{J} \big(   \mathscr{U}^{(\varphi)}, \mathcal{C}^{(\varphi)}_{\mathscr{F}} \big)=  \;\;\;\;\;\;\;\;\;\;\;\;\;\;\;\;\;\;\;\;\;\;\;\;\;\;\;\;\;\;\;\;\;\;\;\;\;\;\;\;\;\;\\ \sum_{i }^{q}\sum_{j }^{k}  \mu^{m,(\varphi)}_{ij}  Tr \Big(s^{(\varphi)}_j \big(log {s^{(\varphi)}_j}-log {c^{(\varphi)}_i }\big)\Big),\varphi \in \{ \ell\}, \\ \sum_{i }^{q}\mu^{m,(\varphi)}_{ij}=1,\;\;\;\;\;\;\;\;\;\;\;\;\;\;\;\;\;\;\;\;\;\;\;\;\;\;\;\;\;\;\;\;\;\;\;\;\;\;\;\\  \ \sum_{j }^{k}  \mu^{m,(\varphi)}_{ij}>0,\;\;\;\;\;\;\;\;\;\;\;\;\;\;\;\;\;\;\;\;\;\;\;\;\;\;\;\;\;\;\;\;\;\;\;\;\;\;\;\\ \;\bigg \lbrace 1-\mathbb{P}r \big \lbrace \mathscr{F}_2(\cdot)=\varnothing \big \rbrace \le \alpha_3 \bigg \rbrace \le \alpha_4,\;\;\;\;\;\;\;\;\;\;\;\;\;\;\;\;\;\;\;\;\;\;\;\;\;
\end{split}
\end{equation*}
where the last constraint indicates that if the player is supposed to lose the game, at least (s)he preserves her possibly winning strategy, with respect to the arbitrary positive scalars $\alpha_i, i\in \{  1, 2, 3, 4 \}$.

\textsc{\textbf{Remark 5.}} \textit{Recall Remark 4 and Definition 2. The agent $\mathscr{P}_2$ is totally aware of
\begin{equation*}
\begin{split}
\; \mathscr{HG}^{(1)}=\langle  \mathscr{G}_1,\mathscr{G}_2 \rangle=\langle  \mathscr{P}_1^{  (Pr \{   \mathscr{P}_2  \})} ,\mathscr{P}_2^{\mathscr{P}_1}        \rangle,
\end{split}
\end{equation*}
whereas the agent $\mathscr{P}_1$ only probabilistically knows $\mathscr{G}_1= \mathscr{P}_1^{  (Pr \{   \mathscr{P}_2  \})}$, that is, the probability of $\mathscr{P}_2$ playing. Consequently, and according to the incomplete information for $\mathscr{P}_1$, (s)he needs to probabilistically define a constraint for herself.}

\textsc{Step 4.} Now, an alternating optimisation based algorithm should be applied. 

This completes the proof.$\; \; \; \blacksquare$

\subsection{Multi-user scenario and a bankruptcy issue}

Let us we have multiple Bobs and Alice owes to $\mathcal{L}$ Bobs, that is, there is a bankruptcy issue. The relative notations are given in Table \ref{table1}. Most importantly, $(\cdot)_{i}, \; I \in \{ 1, 2, 3\}$ is not related to the main game, instead, it hereinafter stands for the bankruptcy issue and the relative $3-$level nested game.

\begin{table}[h]
\begin{center}
\captionof{table}{{List of notations for the bankruptcy issue.}}
 \label{table1} 
\begin{tabular}{p{1.091cm}|p{2.4cm} |p{1.1cm}|p{2.84cm}}
 \hline      
 \hline
Notation        &  Definition  & Notation          &  Definition   \\
\hline
 $\mathcal{P}_b$ &     Bankruptcy prob &     $k$ & Time ($1$st Game)         \\
     $l$ & No. of Decoders    &                    $k^{\prime}$         &          Time ($2$nd Game)           \\
  $\mathcal{K}_b$    &   Bankruptcy time       &  $k^{\prime\prime}$         &          Time ($3$rd Game)                         \\
            $\mu^{(2)}$ &  For 2nd game    & $\mu^{(3)}$                 &      For 3rd game    \\
\hline
 \hline
\end{tabular}
\end{center}
\end{table}

Initially speaking, recall $\mathcal{I} \big(   \mathcal{X},\mathcal{Y} \big) \neq 0$ from \textit{Lemma 1}. LQG-control-cost is a non-linear function of $\mathcal{I}\big(\hat{\mathcal{X}}_l^{( k)};{\mathcal{X}^{( k)}}\big)$ \cite{1, 2, 3, 4}. So, we here write-and-analyse the concave version of it in the following optimisation problem in relation to $\theta^{(\ell)}_{k,1}$ a case in point, where hereinafter the term $(\cdot)_{1}$ is removed for the ease of notation 
\begin{equation*}
\begin{split}
\mathop{{\rm max}}\limits_{\theta_{k,l} \big(  \hat{\mathcal{X}}_l^{( \mathcal{K})}   \big)   } {\rm \; }   \partial_{\Upsilon}^2, \Upsilon=\Upsilon_1 \Bigg( \theta_{k,l}\big(  \hat{\mathcal{X}}_l^{( k)}   \big) ;  dim \bigg(  Null  \Big( \mathcal{I}\big(\hat{\mathcal{X}}_l^{( k)};{\mathcal{X}^{( k)}}\big) \Big) \bigg) \Bigg), \\
s.t. \;\;   \mathcal{X}^{(k+1)}  = \Upsilon_2 \big(    \mathcal{X}^{( k)} ; \theta_{k,l} \big(  \hat{\mathcal{X}}_l^{( k)}   \big) \big)  ,\;\; \;\; \;\; \;\; \;\; \;\; \;\; \;\; \;\; \;\;\;\;\;\;\;
\end{split}
\end{equation*}
aimed at finding the control action $\theta_{k,l}, \forall k \in {  1, \cdots, \mathcal{K}   }, \; \forall l \in {  1, \cdots, \mathcal{L} } $, holding $dim \bigg(  Null  \Big( \mathcal{I}\big(\hat{\mathcal{X}}_l^{( k)};{\mathcal{X}^{( k)}}\big) \Big) \bigg) \longrightarrow \infty$. At the time instant $\mathcal{K}_b$ we experience a bankruptcy situation, that is, according to the any-time-capacity we see 
\begin{equation*}
\begin{split}
\mathcal{P} \left \{    \mathcal{I}  \big(    \mathcal{X}^{(k)};\mathcal{Y}^{(k)}   \big)    > 0 ,  \mathcal{I}  \big(    \mathcal{X}^{(k^{\prime} )};\mathcal{Y}^{(k^{\prime} )}   \big) \ngtr 0  \right \} \ge \mathcal{P}_b, \\ \forall k \in \{   0, \cdots, \mathcal{K}_b-1  \}, k^{\prime}  \in \{   \mathcal{K}_b , \cdots  , \mathcal{K}   \}, \mathcal{K}_b \le \mathcal{K} ,
\end{split}
\end{equation*}
either $\mathcal{K}_b$ including $\infty$ or not $-$ according the any-time-capacity $-$ where $\mathcal{P}_b$ is the bankruptcy probability. 

\subsubsection{First game: For the decoders}
 The first game we have, is among the decoders from a \textit{bankruptcy} based point of view in which Alice owes to $\mathcal{L}$ Bobs and among these Bobs, a coalition is created. In this context, the coalition is $\mathcal{L}_0 \subset \mathcal{L}$ relating to which the transferable-utility pay-off is the LQG controls they are owed. For this game, let us the following parameters be defined: the \textit{transferable}-utility-function is $\mathcal{U}_1 = \mathbb{E} \left \{ \Upsilon \big(k^{\prime} ; \theta_{k^{\prime} ,l_0} \big) \right \}$ where the superscript $(\cdot)_1$ stands for the first game and the expected value is measured over the $\mathcal{L}_0$ Bob set in the coalition. Call $\theta^{*}_{k^{\prime} ,l_0} (\cdot)$ the claimed money by $l_0$-th Bob. According to the \textit{Shapley Value} principle \cite{31} and w.r.t. 
\begin{equation*}
\begin{split}
 \psi(\Omega)=:min \left \{     \theta^{*}_{k^{\prime} ,l_0}      , max \left \{       0,     \theta_{k^{\prime} }  -\theta^{*}_{k^{\prime} ,l_0} \sum_{j \in \mathcal{L}_0 \setminus \{ l_0 \}   }  \theta^{*}_{k^{\prime} ,l_0}  \right \}        \right \},
\end{split}
\end{equation*}
where $\theta_{k^{\prime}} $ is our total budget after the occurrence of the bankruptcy event, the weighted pay-off for every player is obtained by
\begin{equation*}
\begin{split}
\theta_{k^{\prime} ,l_0}=\sum_{l_0 \in \mathcal{L}_0 \subset \mathcal{L}      }  \frac{(||\Omega||-1)!(l_0-||\Omega||)!}{l_0!} \psi(\Omega)-\psi \big (\Omega- \{l_0\} \big).
\end{split}
\end{equation*}
Meanwhile, there seems to be the following conditions to be satisfied \cite{31}: $\sum_{l_0} \theta_{k^{\prime} ,l_0} \nless \theta_{k^{\prime} }$ which indicates the efficiency as well as the fact that there is no chance to create a new coalition; $0 \le \theta_{k^{\prime} ,l_0} \le \theta^{*}_{k^{\prime} ,l_0}$; and $\theta_{k^{\prime} ,l_0} \in \big[          \theta^{(min)}_{k^{\prime} ,l_0}  ,     \theta^{(max)}_{k^{\prime} ,l_0} \big]    $ where the minimum right is $\theta^{(min)}_{k^{\prime} ,l_0}=\psi \big(     \{  l_0  \}    \big)$ $-$ i.e., the individual rationality according to which we can say that no player agrees to receive a less amount of money compared to the case she is out of the coalition $-$ and the maximum right is $\theta^{(max)}_{k^{\prime} ,l_0}=\psi (       \mathcal{L}_0    )-\psi \big(  \mathcal{L}_0    \setminus   \{  l_0  \}    \big)$.

\begin{figure*}
\begin{equation}\label{eq:1}
\begin{split}
\begin{cases}
\; \; \; \; \; \; \; \; \; \; \; \; \; \; \; \; \;\; \; \;\;   \mathop{{\rm \mathbb{I}nf}}\limits_{\theta_{k,l_0} (\cdot)} {\rm \; }   \psi (\Omega ), \\
1.a:\;  \mathcal{U}_2 (k^{\prime}+1)+\mathop{{\rm \mathbb{S}up}}\limits_{\theta_{k^{\prime} ,l_0} (\cdot)} {\rm \; }   \Upsilon (k^{\prime} )+ \partial _{\Upsilon}\mathcal{U}_2 \big(k^{\prime} ;\Upsilon (k^{\prime} ) \big) \sum_{k^{\prime}=\mathcal{K}_b+1}^{\mathcal{K}} \theta_{k^{\prime} ,l_0} \big( k^{\prime} ; \Upsilon (k^{\prime} )  ,\mu^{(2)}_{k^{\prime}}  \big) \mu^{(2)}_{k^{\prime}}   -\bar{\theta}_{k^{\prime} ,l_0} \big( k^{\prime} ; \Upsilon (k^{\prime} )   \big)  +{\sigma}^2 \partial ^2_{\Upsilon\Upsilon} \mathcal{U}_2=0 , \\
1.b:\; \mathop{{\rm max}}\limits_{\theta_{k^{\prime} ,l_0} \big( k^{\prime} ; \Upsilon (k^{\prime} )   \big)} {\rm \; } \left \{ \mathcal{U}_2 = \mathbb{E} \left \{ \sum_{k^{\prime}=\mathcal{K}_b}^{\mathcal{K}}  \Upsilon \big(k^{\prime} ; \theta_{k^{\prime} ,l_0} \big)   \right \} \right \}, \\
1.c:\; {\rm }\Upsilon (k^{\prime} +1)  = \theta_{k^{\prime} ,l_0} \big( k^{\prime} ; \Upsilon (k^{\prime} )   \big)  \pm {\sigma}(k^{\prime} +1),\\
1.d:\;\mathcal{P}_{mf} \Big(\theta_{k^{\prime} +1,l_0} \big(k^{\prime} +1; \Upsilon (k^{\prime}+1 )  \big) \Big)= \sigma^2 (k^{\prime} +1) \partial ^2_{\Upsilon\Upsilon} \left \{ \mathcal{P}_{mf} \Big(\theta_{k^{\prime} ,l_0} \big(k^{\prime} ; \Upsilon (k^{\prime} )  \big) \Big)\right\}  \\\;\; \;\; \;\; \;\; \;\; \;\; \;\; \;\; \;\; \;\; \;\; \;\; \;\; \;\; \;\; \;\; \;\; \;\; \;\; \;\; \;\; \;\; \;\; \;\;\;\;\; \;\;\;\;\;\;\; +\partial _{\Upsilon} \left \{ \mathcal{P}_{mf} \Big(\theta_{k^{\prime} ,l_0} \big(k^{\prime} ; \Upsilon (k^{\prime} )  \big) \Big) \sum_{k^{\prime}=\mathcal{K}_b+1}^{\mathcal{K}}\theta_{k^{\prime} ,l_0} \big( k^{\prime} ; \Upsilon (k^{\prime} )  ,\mu^{(2)}_{k^{\prime} } \big) \mu^{(2)}_{k^{\prime}}   \right \}, \\ 
1.e: \;\;  \mathop{{\rm min}}\limits_{\Omega_1 , \Omega_2}      \underbrace{\mathcal{P}_{mf} \Big(\theta_{k^{\prime}_b,l_0} \big(k^{\prime}_b; \Upsilon (k^{\prime}_b)  \big) \Big)}_{\Theta(\Phi_1)}   \left |  \right|        \underbrace{\mathcal{P}_{mf} \Big(\theta_{\mathcal{K}_b,l_0} \big(\mathcal{K}_b; \Upsilon (\mathcal{K}_b)  \big) \Big)}_{\Theta(\Phi_2)} .
\end{cases}
\end{split}
\end{equation}
\end{figure*}

\begin{figure*}
\begin{equation}\label{eq:2}
\begin{split}
\begin{cases}
\; \; \; \; \; \; \; \; \; \; \; \; \; \; \; \; \;\; \; \;\;   \mathop{{\rm \mathbb{I}nf}}\limits_{\theta_{k,l_0} (\cdot)} {\rm \; }   \psi (\Omega ), s.t.\; 1.a-1.d \; as\;2.a-2.d;\; and 2.e:\;  \partial_{k\prime \prime}\mathcal{U}_3 +\mathop{{\rm \mathbb{S}up}}\limits_{\theta_{k^{\prime\prime} ,l_0} (\cdot)} {\rm \; }    \varphi (k^{\prime \prime} ) \\
\; \; \; \; \; \; \; \; \; \; \; \; \; \; \; \; \; \; \; \; + \partial _{ \varphi}\mathcal{U}_3 \big(k^{\prime \prime} ; \varphi (k^{\prime \prime} ) \big) \int_{\tau^{\prime}} \theta_{k^{\prime \prime} ,l_0} \big( k^{\prime \prime} ;  \varphi (k^{\prime \prime} )  ,\mu^{(3)}_{k^{\prime \prime}}  ; \tau^{\prime} \big) \mu^{(3)}_{k^{\prime \prime}} (\tau^{\prime}) d\tau^{\prime}   -\bar{\theta}_{k^{\prime \prime} ,l_0} \big( k^{\prime \prime} ; \varphi  (k^{\prime \prime} )   \big)  +{\sigma}^2 \partial ^2_{\varphi \varphi } \mathcal{U}_3=0 , \\
2.f:\; \mathop{{\rm min}}\limits_{\theta_{k^{\prime\prime} ,l_0} \big( k^{\prime\prime} ; \varphi  (\cdot )   \big)} {\rm \; } \left \{ \mathcal{U}^{(\tau,\tau^{\prime})}_3 = \mathbb{E} \left \{ \int_{\mathcal{K}_b}^{\mathcal{K}}  \left | \varphi  \big(\tau, k^{\prime\prime} ; \theta_{k^{\prime\prime} ,l_0} \big)-\varphi  \big(\tau^{\prime}, k^{\prime\prime} ; \theta_{k^{\prime\prime} ,l_0} \big) \right | dk^{\prime\prime}  \right \} \right \}, \forall \tau^{\prime} \in \{ 1,\cdots,N \} \setminus \{ \tau \} ,\\
2.g:\; {\rm }\partial_{k\prime \prime}\varphi  (\tau,k^{\prime\prime} )  = \omega \big( \tau, \theta_{k^{\prime\prime} ,l_0} ,  k^{\prime\prime} \big)  \pm {\sigma},\\
2.h:\;\partial_{k\prime \prime}\mathcal{P}_{df} \Big(\theta_{k^{\prime\prime} ,l_0} \big(k^{\prime\prime} ; \varphi  (k^{\prime\prime} )  \big) \Big)= \sigma^2 \partial ^2_{\varphi \varphi } \left \{ \mathcal{P}_{df} \Big(\theta_{k^{\prime\prime} ,l_0} \big(k^{\prime\prime} ; \varphi  (k^{\prime\prime} )  \big) \Big)\right\}  \\\;\; \;\; \;\; \;\; \;\; \;\; \;\; \;\; \;\; \;\; \;\; \;\;\; \;\; \;\; \;\; \;\; \;\; \;\; \;\; \;\; \;\; \;\; \;\; \;\; \;\;  +\partial _{\varphi } \left \{ \mathcal{P}_{df} \Big(\theta_{k^{\prime\prime} ,l_0} \big(k^{\prime\prime} ; \varphi  (k^{\prime\prime} )  \big) \Big) \int_{\tau^{\prime}}\theta_{k^{\prime\prime} ,l_0} \big( k^{\prime\prime} ; \varphi  (k^{\prime\prime} )  ,\mu_{k^{\prime\prime} } ; \tau^{\prime}\big) \mu^{(3)}_{k^{\prime\prime}} (\tau^{\prime}) d\tau^{\prime} \right \}.
\end{cases}
\end{split}
\end{equation}
\end{figure*}
\subsubsection{Second game: For time instants}
For the second game we experience, the players are assigned to the LQG control signal vector for every individual Bob over the time interval $[\mathcal{K}_b,\mathcal{K}]$. In this scenario which is a discrete mean-field-game (MFG), we have an $N+1$-player \textit{mean-field-limit} based game \cite{32} for which the sensitivity about 1 major player, that is, the initial condition $\mathcal{K}_b$, should be considered by other players. For this game, let us the following parameters be defined: $\forall l_0 \in \{   0, \cdots, \mathcal{L}_0  \}$, the utility-function is $\mathcal{U}_2 = \mathbb{E} \left \{ \sum_{k^{\prime}=\mathcal{K}_b}^{\mathcal{K}}  \Upsilon \big(k^{\prime} ; \theta_{k^{\prime} ,l_0} \big)  \right \}$ where the superscript $(\cdot)_2$ stands for the second game; for $\theta_{k^{\prime} ,l_0} \big(k^{\prime} ; \Upsilon (k^{\prime} )  \big) $, the PMF is $\mathcal{P}_{mf} \Big(\theta_{k^{\prime} ,l_0} \big(k^{\prime} ; \Upsilon (k^{\prime} )  \big) \Big)  \ge 0$ where $ \left | \sum_{k^{\prime}=\mathcal{K}_b}^{\mathcal{K}} \mathcal{P}_{mf} \Big(\theta_{k^{\prime} ,l_0} \big(k^{\prime} ; \Upsilon (k^{\prime} )  \big) \Big)  \right |=1$; the average is $\bar{\theta}_{k^{\prime} ,l_0} \big(k^{\prime} ; \Upsilon (k^{\prime} )  \big)=\sum_{k^{\prime}=\mathcal{K}_b}^{\mathcal{K}} \theta_{k^{\prime} ,l_0} \big(k^{\prime} ; \Upsilon (k^{\prime} )  \big) \mathcal{P}_{mf} \Big(\theta_{k^{\prime} ,l_0} \big(k^{\prime} ; \Upsilon (k^{\prime} )  \big) \Big)  $; and $\sigma$ is a noise. 

\begin{theorem} \label{T1} \textit{Our Bi-level game has the solution as in Eq. \eqref{eq:1} where $\mu^{(2)}_{k^{\prime}} $ is a PDF relating to the interactions with the major-player $\mathcal{K}_b$\footnote{Something that, in parallel with $\mathcal{P}_{mf} \Big(\theta_{k^{\prime} ,l_0} \big(k^{\prime} ; \Upsilon (k^{\prime} )  \big) \Big) $, indicates that the PMF of the interactions among the agents in our second game is 2-Dimensional.} where $ k^{\prime}_b \in \left \{ \underbrace{[\mathcal{K}_b,\mathcal{K}]}_{k^{\prime}} \setminus \{\mathcal{K}_b\} \right \} $, i.e., $k^{\prime}_b \in ( \mathcal{K}_b,\mathcal{K}] $.}\end{theorem}

\textbf{\textsc{Proof:}} In order to understand how to right the relative mathematical equations, that is, the \text{Hamilton-Jacobi-Bellman} one $-$ which finds the local interactions $-$ and the \textit{Fokker-Planck-Kolmogorov} one $-$ which finds the PDF of the total interactions $-$ revising them to the \textit{McKean-Vlasov} based ones, please see e.g. \cite{32}. Indeed, ADMM is supposed to decrease the divergence of the 2 conditions while finding one of them fixing another one and vice versa. Finally speaking, $\mu^{(2)}_{k^{\prime}} $ is the PMF w.r.t. the major agent from others' side.

This completes the proof.$\; \; \; \blacksquare$

\subsubsection{Third game: For phase transitions}

 We are fully aware of the fact that PMFs are mapped into the bifurcation of phases, i.e., phase transitions \cite{33, 34}. Consequently, one can apply Kuramoto-model instead of $1.e$ in \textit{Theorem 1} which results in the following theorem. Prior of this, for this game, let us the following parameters be defined: the utility-function is $\mathcal{U}_3$ where the superscript $(\cdot)_3$ stands for the third game; for $\theta_{k^{\prime\prime} ,l_0} \big(k^{\prime\prime} ; \varphi (k^{\prime\prime} )  \big) $, the probability distribution function (PDF) is $\mathcal{P}_{df} \Big(\theta_{k^{\prime\prime} ,l_0} \big(k^{\prime\prime} ; \varphi (k^{\prime\prime} )  \big) \Big)  \ge 0$ where $ \left | \int_{k^{\prime\prime}=\mathcal{K}_b}^{\mathcal{K}} \mathcal{P}_{df} \Big(\theta_{k^{\prime\prime} ,l_0} \big(k^{\prime\prime} ; \varphi (k^{\prime\prime} )  \big) \Big)  \right |=1$; the average is $\bar{\theta}_{k^{\prime\prime} ,l_0} \big(k^{\prime\prime} ; \varphi (k^{\prime\prime} )  \big)=\int_{k^{\prime\prime}=\mathcal{K}_b}^{\mathcal{K}} \theta_{k^{\prime\prime} ,l_0} \big(k^{\prime\prime} ; \varphi (k^{\prime\prime} )  \big) \mathcal{P}_{df} \Big(\theta_{k^{\prime\prime} ,l_0} \big(k^{\prime\prime} ; \varphi (k^{\prime\prime} )  \big) \Big)  $; and $\sigma$ is a noise.  

\begin{theorem} \label{T2} \textit{Our Bi-level game can be revised into a 3-level nested game by modifying $1.e$ as in Eq. \eqref{eq:2}. $\partial_{k^{\prime}} \varphi_{\tau} = \big( \omega_{\tau} +u_{\tau} (k^{\prime}) \big )  +\partial_{k^{\prime}} \zeta_{\tau},   k^{\prime}= \mathcal{K}_b,\cdots,\mathcal{K}, \varphi \in [0,2 \pi], \tau \in [1, \infty), u_{\tau}(k^{\prime})=\frac{D}{N}\sum_{\tau^{\prime}=1,\cdots,N}sin\big(  \varphi_{\tau^{\prime}}  -\varphi_{\tau}   \big)$ and $D \in \mathbb{R}$ is the coupling gain.}\end{theorem}

\textsc{\textbf{Proof:}} Although $k^{\prime\prime}$ the same as $k^{\prime}$ belongs to $\mathcal{K}_b,\mathcal{K}$, by the way the reason that we respectively define a discrete MFG and a continuous one for respectively the second and the third games are justified as follows. The control law according to which the second game is actualised is a discrete time process $-$ control actions $-$ , inversely, the control law w.r.t. which the third game is realised is a continuous time process $-$ i.e., phase transitions. In order to understand what \textit{Kuramoto} model is, please see e.g. \cite{35}. Finally speaking, $\mu^{(3)}_{k^{\prime\prime}} (\tau^{\prime})  $ is the PMF w.r.t. the major agent from others' side. 

This completes the proof.$\; \; \; \blacksquare$

%
%

\begin{theorem} \label{T3} \textit{Our framework is \textit{not completely controllable but stabilisable.} }\end{theorem} 

\textsc{\textbf{Proof:}} Although in our framework some decoders may experience information leakage, by the way the overall system rate can be non-zero \cite{1, 2}. Please see e.g. \cite{13} in order to understand how a system is \textit{not completely controllable but stabilisable}. This kind of interpretation originally comes from $(0, \nu_1,\nu_2)-$stabilisability principle\footnote{See e.g. \cite{36} to understand what it is.} where stabilisability is satisfied in an annulus w.r.t. the 2 concentric balls of different radii $\nu_1$ and $\nu_2$ centred at the origin. Meanwhile, $\nu_1$ and $\nu_2$ are respectively the lower-bound and the upper-bound of the stabilisability-region. In other words, $\nu_1$ is sensitive to the information leakage, while $\nu_2$ is satisfied by $\mathcal{P} \left \{ \mathcal{I} \big( \mathcal{X}^{(k)};\mathcal{Y}^{(k)} \big) > 0 , \mathcal{I} \big( \mathcal{X}^{(k^{\prime} )};\mathcal{Y}^{(k^{\prime} )} \big) \ngtr 0 \right \} < \mathcal{P}_b$. 

This completes the proof.$\; \; \; \blacksquare$

\section{Numerical results}

We have done our simulations w.r.t. the Bernoulli-distributed data-sets using GNU Octave of version $4.2.2$
on Ubuntu $16.04$.

Initially speaking, let us also define the greedy Algorithm \ref{f}. Fig. \ref{F3} compares Algorithms \ref{ffff} and \ref{f}. 

\begin{algorithm}
\caption{\textcolor{black}{A greedy algorithm to $\mathscr{P}_1$. }}\label{f}
\begin{algorithmic}
\STATE \textbf{\textsc{Initialisation.}} 

\textbf{while} $ \mathbb{TRUE}$ \textbf{do} Iteratively find the solution. \textbf{endwhile} 

\textbf{end}

\textbf{\textsc{Output:}} $\mathcal{P}\big( \mathcal{Y}|\mathcal{X} \big)$
\end{algorithmic}
\end{algorithm}

The first sub-figure demonstrates the cumulative distribution function (CDF) of the iterations needed for our proposed Algorithms \ref{ffff} and \ref{f} to be converged while changing $\alpha$ for Algorithm \ref{ffff}. The degradation of the performance while decreasing $\alpha$ is totally obvious. It is also revealed that Algorithm \ref{ffff} out-performs the greedy Algorithm \ref{f}. Moreover, a partially less acceptable performance is observed while decreasing $\alpha$ which, as discussed, shows dissipativity.

 The second sub-figure demonstrates $\mathcal{I}\big(   \mathcal{S},\mathcal{Y} \big)$ versus the normalised version of $\mathcal{I}\big(   \mathcal{X},\mathcal{Y} \big)$ while changing $\alpha$ for Algorithm \ref{ffff}. It is again revealed that the Algorithm \ref{ffff} out-performs the greedy Algorithm \ref{f}. Again, a partially more favourable performance is seen for the higher value of the term $\alpha$. 

Finally, the third sub-figure shows the loss in terms of percentage against the normalised version of the iteration regime while changing $\alpha$ for Algorithm \ref{ffff}. It is again revealed that the Algorithm \ref{ffff} has a significantly more adequate performance in comparison with the greedy Algorithm \ref{f}. Additionally, the influence of a change in the value of $\alpha$ on the overall performance of the system is obvious.

\begin{figure}[t]
\centering
\subfloat{\includegraphics[trim={{33 mm} {84 mm} {21 mm} {90mm}},clip,scale=0.574]{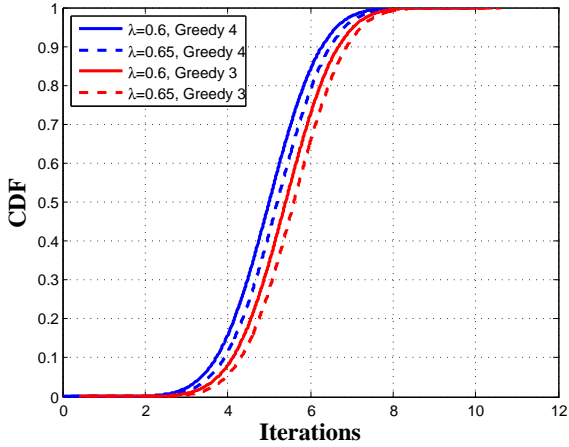}} 
\caption{CDF of the iterations required for the convergence of Algorithms \ref{fffff} and \ref{ffffff} changing $\lambda$ while considering $\mathcal{K}_b=6 \; sec$, $\mathcal{K}=86400 \; sec$, i.e., $24$ hours, and $\mathcal{L}=9$ and $\mathcal{L}_0=5$. } \label{F13}
\label{fig:EcUND} 
\end{figure}

Defining the dissatisfaction-rate $\lambda=1-\frac{\theta_{k,l_0}}{\theta^{*}_{k,l_0}}$, Fig. \ref{F13} shows CDF of the iterations needed for our proposed Algorithms \ref{fffff} and \ref{ffffff} to be converged. The degradation of the performance while increasing $\lambda$ is totally obvious. Meanwhile, it is revealed that our proposed algorithm \ref{ffffff} out-performs the Algorithm \ref{fffff}. 

%

\textsc{\textbf{Remark 6:}} The reason of the out-performance of our Kuramoto based Algorithm \ref{ffff} over the ADMM based Algorithm \ref{fffff} can be technically interpreted as follows. Theoretically, ADMM needs a huge search region, consequently, it needs more time to blindly search for a global equilibrium in order to find a more accurate globality. Inversely, the Kuramoto model is supposed to intelligently find an appropriate global equilibrium. In other words, the ADMM should be able to move over a larger search region to find the global optima not considering the fact that the major player has an extremely tiny sensitivity to the minor ones, inversely, only the sensitivity of the minor ones as the \textit{Followers} are considered in the Kuramoto model since the major player is the \textit{Leader}. This is the tremendously important superiority of the Kuramoto model in comparison with the ADMM in our nested game-theoretical solution. Let us terminate our discussion with the following exemplification. Consider that a father has 2 children. According to ADMM, the father should blindly find an equilibrium between them. Conversely, according to the Kuramoto model, he has to smartly carefully take into account the child who has a genetic jump and he/she is supposed to be more effective for the future. Indeed, in the Kuramoto model $-$ initially suggested for biological oscillations where each oscillator is connected to another one $-$ the interactions can be limited over a certain graph although the topology is a complete graph \cite{37, 38}.

\textsc{\textbf{Remark 7:}} Algorithms \ref{fffff} and \ref{ffffff} are of the complexity-orders of respectively $\mathcal{O} \big (3N+N \log N +N^2 \big)$ and $\mathcal{O} \big (2N+N \log N +N^2 \big)$.

\begin{figure}[t]
\centering
\subfloat{\includegraphics[trim={{13 mm} {71 mm} {11 mm} {70mm}},clip,scale=0.46]{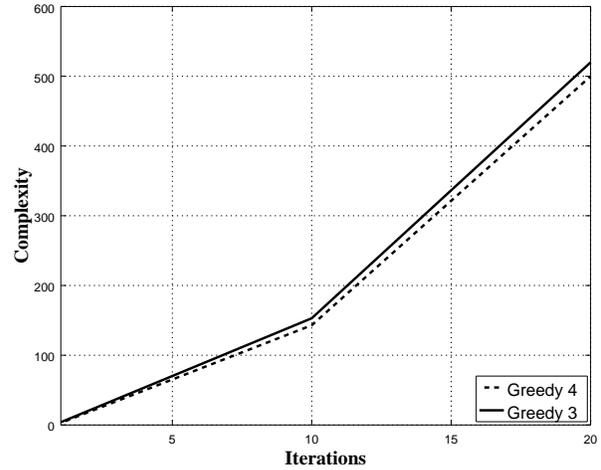}} 
\caption{\textcolor{black}{Complexity of Algorithms \ref{fffff} and \ref{ffffff} in terms of $\mathcal{O}(\cdot)$ vs. the iteration regime $-$ \textit{Remark 7}.} } \label{F14}
\label{fig:EcUND} 
\end{figure}

\textcolor{black}{Fig. \ref{F14} also demonstrates the discussion given in \textit{Remark 7} in terms of $\mathcal{O}(\cdot)$.}

\begin{algorithm}
\caption{\textcolor{black}{An algorithm to find $\theta_{k,l_0}$: Theorem 1.}}\label{fffff}
\begin{algorithmic}
\STATE \textbf{\textsc{Initialisation.}}

$\;\;$$\;\;\;\;\;\;\;\;$ \textbf{while} $\mathbb{TRUE}$ \textbf{do}  Solve $(1)$  \textbf{endwhile} 

\textbf{end}

\end{algorithmic}
\end{algorithm}

\begin{algorithm}
\caption{\textcolor{black}{A greedy algorithm to find $\theta_{k,l_0}$: Theorem 2.}}\label{ffffff}
\begin{algorithmic}
\STATE \textbf{\textsc{Initialisation.}}

$\;\;$$\;\;\;\;\;\;\;\;$ \textbf{while} $\mathbb{TRUE}$ \textbf{do}  Solve $(2)$  \textbf{endwhile} 

\textbf{end}

\end{algorithmic}
\end{algorithm}

\section{conclusion}
The privacy-utility trade-off for the problem of control over noisy communication channels was evaluated in this paper. The performance of the network from a joint continuous MFG-theoretical point of view was also explored. In relation to our MFG, the PDF of the population density was two-dimensional. The control-laws were generalised from a dissipativity theoretic standpoint. Then, a Blahut-Arimoto algorithm was proposed according to the federated learning principle in order to find the PMFs for the privacy-utility trade-off. We investigated this from a time-varying graph theoretic perspective. We derived some novel results in the
context of new theorems and propositions. Finally, simulations showed us that our proposed algorithm has a remarkably more adequate performance compared to a greedy algorithm. More interestingly, simulations showed us the effect of a change in the quantity of $\alpha$ according to the dissipativity theory.

\markboth{IEEE, VOL. XX, NO. XX, X 2021}%
{Shell \MakeLowercase{\textit{et al.}}: Bare Demo of IEEEtran.cls for Computer Society Journals}

\begin{thebibliography}{}
\bibitem{a1}Z. Yuhua, J. Xiangdong, W. Yuxin, C. Shengnan, ''A Research on Age of Information Minimization Scheme of Wireless Sensor Network Assisted by UAV,'' \textit{6th Int. Conf. Intell. Comput. Sig. Proc. (ICSP)}, Xi'an, China, 2021.
\bibitem{a2}F. Wu, H. Zhang, J. Wu, L. Song, Z. Han, H. V. Poor, ''UAV-to-Device Underlay Communications: Age of Information Minimization by Multi-agent Deep Reinforcement Learning,'' https://arxiv.org/abs/2003.05830, 2020.
\bibitem{a3}S. Wang, M. Chen, Z. Yang, C. Yin, W. Saad, S. Cui, H. V. Poor, ''Reinforcement Learning for Minimizing Age of Information in Real-time Internet of Things Systems with Realistic Physical Dynamics,'' https://arxiv.org/abs/2104.01527, 2021.
\bibitem{a4}J. Hu, H. Zhang, L. Song, R. Schober, H. V. Poor,'' Cooperative Internet of UAVs: Distributed Trajectory Design by Multi-agent Deep Reinforcement Learning,'' https://arxiv.org/abs/2007.14297, 2021. 
\bibitem{a5}J. Hu, H. Zhang, L. Song, Z. Han, H. V. Poor, ''Reinforcement Learning for a Cellular Internet of UAVs: Protocol Design, Trajectory Control, and Resource Management,'' https://arxiv.org/abs/1911.08771, 2019. 
\bibitem{a6}
S. Zhang; H. Zhang, Z. Han, H. V. Poor, L. Song, ''Age of Information in a Cellular Internet of UAVs: Sensing and Communication Trade-Off Design,'' \textit{IEEE Trans. Wireless Commun.}, Vol. 19, no. 10, pp. 6578-6592, 2020.
\bibitem{1}S. Tatikonda, S. Mitter,"Control over noisy channels," \textit{IEEE Trans. Automatic Control}, Vol. 49, no. 7, pp. 1196-1201, 2004.
\bibitem{2}M. Zamanipour, "A Novelty in \textit{Blahut-Arimoto} Type Algorithms: Optimal Control over Noisy Communication Channels," \textit{IEEE Trans. Vehicular Technol} Vol. 69, no. 6, pp. 6348-6358, 2020.
\bibitem{3}A. Khina, E. R. Garding, G. M. Pettersson, V. Kostina, B. Hassibi, ''Control Over Gaussian Channels With and Without Source-Channel Separation,''  \textit{IEEE Trans. Auto. Control} Vol. 64, no. 9, pp. 3690 2019. 
\bibitem{4}V. Kostina, B. Hassibi, ''Fundamental limits of distributed tracking,''  https://arxiv.org/abs/1910.02534, 2020. 
\bibitem{5}C. Cai, S. Verdu, ''Conditional Renyi divergence saddlepoint and the maximization of $\alpha$-mutual information,'' \textit{Entropy}, Vol. 21, no. 10, 2019.
\bibitem{6}M. Zamanipour, "A Novel \& Stable Stochastic-Mean-field-Game for Lossy Source-coding Paradigms: A Many-Body-Theoretic Perspective," \textit{IEEE ACCESS}, Vol. 7, pp. 111355-111362, 2019. 
\bibitem{7}M. Zamanipour, "Fast-and-Secure State-Estimation in Dynamic-Control over Communication Channels: A Game-theoretical Viewpoint," \textit{IEEE Trans. Sig. Info. Process. Nets.}, Vol. PP, no. 99, pp. 1-1, 2020. 
\bibitem{8}M. Xiao, W. Zheng, G. Jiang, J. Cao, ''Stability and Bifurcation of Delayed Fractional-Order Dual Congestion Control Algorithms,'' \textit{IEEE Trans. Auto. Control}, Vol. 62, no. 9, pp. 4819-4826, 2017.
\bibitem{9}Y. Liu, H. Su, ''Necessary and Sufficient Conditions for Containment in Fractional-Order Multiagent Systems via Sampled Data,'' \textit{IEEE Trans. Sys. Man. Cyber. Systems}, Vol. PP, no. 99, pp. 1-1, 2020.
\bibitem{10}R. Fox, R. Shin, W. Paul, Y. Zou, D. Song, K. Goldberg, P. Abbeel, I. Stoica, ''Hierarchical Variational Imitation Learning of Control Programs,'', https://arxiv.org/abs/1912.12612, 2019.
\bibitem{11}B. G. Pegueroles, G. Russo, ''Sensitivity and safety of fully probabilistic control,'' https://arxiv.org/abs/1903.09484, 2019. 
\bibitem{12}S. Salimi, S. Dey and A. Ahlen, ''Sequential Detection of Deception Attacks in Networked Control Systems with Watermarking,'' \textit{IEEE 18th Euro. Control Conf. (ECC)}, Italy, June 2019.
\bibitem{13}A. Makhdoumi, S. Salamatian, N. Fawaz, M. Medard, ''From the Information Bottleneck to the Privacy Funnel,'' \textit{IEEE Info. Theory Workshop (ITW)}, Australia, Nov 2014.
\bibitem{14}P. S. Maybeck, "\textit{Stochastic Models, Estimation and Control (I and II)}", Academic Press, New York, 1982.
\bibitem{15}A. El Gamal and Y.-H. Kim, ''\textit{Network information theory.}'' Cambridge
University Press, 2011.
\bibitem{16}X. Chen, M. Huang, ''Linear-quadratic mean field control: The invariant subspace method,'' \textit{Automatica}, Vol. 107, pp. 582-586, 2019.
\bibitem{17}Q. Wang, S. R. Kulkarni, and S. Verdu, ''Divergence Estimation of Continuous Distributions
Based on Data-Dependent Partitions,'' \textit{IEEE Trans. Info. Theory,}, Vol. 51, no. 9, pp. 3064-3074, 2005. 
\bibitem{18}P. S. Maybeck, "\textit{Stochastic Models, Estimation and Control (I and II)}",
Academic Press, New York, 1982.
\bibitem{19}A. Kulkarni, J Fu, ''Opportunistic Synthesis in Reactive Games under Information Asymmetry,'' https://arxiv.org/abs/1906.05847, 2019.
\bibitem{20}C. Bakker, A. Bhattacharya, S. Chatterjee, D. L. Vrabie, ''Hypergames and Cyber-Physical Security for Control Systems,'' https://arxiv.org/abs/1809.02240, 2019.
\bibitem{21}Abhishek N. Kulkarni; Huan Luo; Nandi O. Leslie; Charles A. Kamhoua; Jie Fu, ''Deceptive Labeling: Hypergames on Graphs for Stealthy Deception,'' \textit{IEEE Control Sys. Letters,} Vol. 5, no. 3, 2020.
\bibitem{22}L. A. Zadeh, ''Fuzzy sets,'' \textit{Info. Control}, Vol. 8, pp. 338-353, 1965.
\bibitem{31} R. B. Myerson, ''\textit{Game theory,}'' Harvard university press., 2013.
\bibitem{32}M. Duerinckx, L. Saint-Raymond, ”Lenard-Balescu correction to meanfield theory,” https://arxiv.org/abs/1911.10151, 2019.
 \bibitem{33}K. Crammer, P. P. Talukdar, F. C. N. Pereira, ''A Rate-Distortion One-Class Model and its Applications to
Clustering,'' \textit{ICML '08: Proceedings. 25th int. conf. Machine learning}, July 2008.
\bibitem{34}K. Rose, ''A Mapping Approach to Rate-Distortion
Computation and Analysis,'' \textit{IEEE Trans. Info. Theory}, Vol. 40, no. 6, pp. 1939-1952, 1994.
\bibitem{35}H. Yin, P. G. Mehta, S. P. Meyn, U. V. Shanbhag, ''Synchronization of Coupled Oscillators is a Game,'' \textit{IEEE Trans. Automatic Control}, Vol. 57, no. 4, pp. 920-935, 2012. 
\bibitem{36}L. Ma, W. Zhang, Y. Zhao, ''Study on stability and stabilizability of discrete-time
mean-field stochastic systems,'' \textit{J. Franklin Inst.}, Vol. 356, no. 4, pp. 2153-2171, 2019.

\bibitem{37}A. Jadbabaie, N. Motee, and M. Barahona, “On the stability of the Kuramoto model of coupled nonlinear oscillators,” in American Control Conference, Proc. 2004, vol. 5, pp. 4296–4301, IEEE, 2004.
\bibitem{38}S. H. Mousavi, ‘’Problems in Control, Estimation, and Learning in Complex Robotic Systems,’’ \textit{Lehigh University}, 2019.
\end{thebibliography}
\end{document}